\documentclass{aa}
\usepackage{amsmath}
\usepackage{txfonts}
\usepackage{braket}
\usepackage{amssymb}
\usepackage{enumerate}
\usepackage{graphicx}
\usepackage{subcaption}
\usepackage{gensymb}
\usepackage{wasysym}

\begin{document}

\title{PORTAL--three-dimensional polarized (sub)millimeter line radiative transfer}
\titlerunning{PORTAL--polarized line radiative transfer}
\author{Boy Lankhaar
        \and
        Wouter Vlemmings
        }
\institute{Department of Space, Earth and Environment, Chalmers University of Technology, Onsala Space Observatory, 439 92 Onsala, Sweden \\
\email{boy.lankhaar@chalmers.se}}

\date{Received ... ; accepted ...}

\abstract
   {Magnetic fields are important to the dynamics of many astrophysical processes and can typically be studied through polarization observations. Polarimetric interferometry capabilities of modern (sub)millimeter telescope facilities have made it possible to obtain detailed velocity resolved maps of molecular line polarization. To properly analyze these for the information they carry regarding the magnetic field, the development of adaptive three-dimensional polarized line radiative transfer models is necessary.}
   {We aim to develop an easy-to-use program to simulate the polarization maps of molecular and atomic (sub)millimeter lines in magnetized astrophysical regions, such as protostellar disks, circumstellar envelopes, or molecular clouds.}
   {By considering the local anisotropy of the radiation field as the only alignment mechanism, we can model the alignment of molecular or atomic species inside a regular line radiative transfer simulation by only making use of the converged output of this simulation. Calculations of the aligned molecular or atomic states can subsequently be used to ray trace the polarized maps of the three-dimensional simulation.}
   {We present a three-dimensional radiative transfer code, POlarized Radiative Transfer Adapted to Lines (PORTAL), that can simulate the emergence of polarization in line emission through a magnetic field of arbitrary morphology. Our model can be used in stand-alone mode, assuming LTE excitation, but it is best used when processing the output of regular three-dimensional (nonpolarized) line radiative transfer modeling codes. We present the spectral polarization map of test cases of a collapsing sphere and protoplanetary disk for multiple three-dimensional magnetic field morphologies.}
   {}
   {}


\maketitle

\section{Introduction}
Magnetic fields permeate the Universe and often play an important role in the dynamics of astrophysical processes \citep{crutcher:12, vlemmings:13, crutcher:19}. It is difficult to directly observe magnetic fields; one typically has to use the polarization properties of the observed light \citep[e.g.,][]{han:17}. For (sub)millimeter interferometers, such as ALMA, magnetic field detection is done mostly through dust \citep[e.g.,][]{hull:17} and line polarization observations \citep[e.g.,][]{vlemmings:17}. However, it has recently become increasingly clear that dust polarization does not always faithfully trace the magnetic field morphology, but instead it can be affected by processes such as self-scattering \citep{kataoka:15, kataoka:17}. Line polarization observations are not affected by such processes, and therefore they likely trace the magnetic field structure of the observed region. However, to interpret line polarization observations, modelers have to defer to the theory of \citet{goldreich:81}, which relies on the large velocity gradient (LVG) approximation, and therefore they cannot treat three-dimensional (3D, magnetic field) structures. 

In this paper, we present POlarized Radiative Transfer Adapted to Lines (PORTAL)\footnote{The source code of PORTAL is available on GitHub at \url{https://github.com/blankhaar/PORTAL}.}, which is a 3D polarized radiative transfer code that simulates the emergence of polarization in the emission of atomic or molecular (sub)millimeter lines. PORTAL can be used in stand-alone mode or process the output of regular 3D radiative transfer codes. We are able to model the emergence of linear polarization in (sub)millimeter lines through two main approximations: (i) the strong magnetic field approximation and (ii) the anisotropic intensity approximation.\ We show that both of them are valid in the majority of astrophysical regions. 

Regular radiative transfer models of astrophysical environments only take the total radiation intensity and its effect on the local isotropic populations into account \citep{vandertak:07, brinch:10}. The local populations are determined by the balance of collisional and radiative events that both excite and de-excite the populations of the molecular and atomic species. Collisional events are isotropic and a function of the density and temperature of the environment. At the outset, for unaligned quantum states, spontaneous emission events are also isotropic, that is,~the direction of the next spontaneously emitted photon of a certain molecule is random. The probability of absorption of those randomly directed photons, however, need not be isotropic \citep{goldreich:81}. Anisotropy in the local absorption of photons aligns the quantum states that are associated with the line-transition, which in turn leads to polarization in the emission \citep{morris:85, landi:06}. By considering the directional dependance of the photon-escape probability in a medium with an anisotropic velocity gradient, \citet{goldreich:81} showed that radiation emitted from such a system is partially polarized. This effect is known in the literature as the Goldreich-Kylafis (GK) effect. It is strongest for lines with optical depth around unity in regions where the collisional rates are not so high as to quench the molecular or atomic alignment. Provided the magnetic field precession rate is 10-100 times stronger than radiative and collisional rates, which we show to be the case in most astrophysical regions in Section \ref{sec:comp_align}, the line polarization traces the magnetic field projected onto the plane of the sky with a $90^{\circ}$ ambiguity. 

Numerical modeling of the GK effect has been based on the theory presented in \citet{goldreich:82}. In such models, the perpendicular and parallel components (with respect to the projected magnetic field direction) of the radiation field are propagated through a medium with an anisotropic velocity gradient. The velocity-gradient is so strong that the large velocity gradient (LVG) approximation can be employed. The LVG escape probability is a function of the velocity-gradient and is therefore anisotropic. This leads to alignment in the molecular or atomic states associated with the transition under investigation. Because of this, the emitted radiation is partially polarized. \citet{deguchi:84} later showed that in order to accurately model the GK effect, it is vital to perform comprehensive (polarized) excitation modeling of the molecular or atomic quantum states and also of the ones that are not associated with the transition under investigation. \citet{cortes:05} showed that an external anisotropic radiation source, such as a nearby stellar object, can enhance the polarized emission significantly. These numerical models only considered the one-dimensional propagation of polarized radiation, and the representation of the radiation field in perpendicular and parallel components is only valid when the magnetic field direction is constant over the investigated path. Furthermore, because of its heavy reliance on the LVG approximation, numerical modeling based on \citet{goldreich:82} can only consider the introduction of anisotropy in the escape probability through an anisotropic velocity gradient. In light of recently developed polarimetric capabilities of interferometers, such as ALMA, these types of approximations cannot be afforded anymore. Rather, one needs comprehensive modeling of the 3D radiative transfer and its anisotropy, taking both the spatial and velocity structure into account for the astrophysical region under investigation as well as the 3D structure of the magnetic field. 

In this paper, we demonstrate how such modeling can be attained. By using two (main) approximations, we show that regular (nonpolarized) radiative transfer codes can be extended with polarization capabilities. In Section 2, we introduce these approximations and show their simplifying impact on the theory of line polarization. In Section 3, we show how our PORTAL code provides the option of computing the emerging polarization using the output from a regular 3D radiative transfer code, in particular LIME \citep{brinch:10}. In Section 4, we present the capabilities of PORTAL through the simulation of the emergence of polarization in a protoplanetary disk and a collapsing sphere. We discuss our results in Section 5 and conclude in Section 6.

\section{Theory}
We describe the introduction of anisotropy in the molecular or atomic populations through an anisotropic radiation field using the formalism of \citet{landi:06}. We make the following approximations: 

First, we assume the magnetic field precession rate is way higher than collisional and radiative rates. We call this the strong magnetic field approximation. The magnetic precession rate is in the order of s$^{-1}$/mG for diamagnetic (i.e.,~weakly magnetizable) molecules. Typical collisional rates are on the order of $10^{-5} \left(\frac{n_{H_2}}{10^6 \ \mathrm{cm}^{-3}}\right)\ \mathrm{s}^{-1}$ and radiative rates are on the order of $10^{-4} \ \mathrm{s}^{-1}$ for a transition at $100$ GHz with a dipole moment of $0.1$ Debye, which is shone upon isotropically by $400$ Kelvin black-body radiation. Therefore, for almost all molecules, magnetic field interactions already dominate at very weak magnetic fields ($\mu$G). Under the assumption of a strong magnetic field, many terms in the polarized density-equations can be dropped \citep{landi:06}. The strong magnetic field approximation is also invoked by \citet{goldreich:81}. In Section \ref{sec:comp_align} we discuss special cases where a dominant magnetic field cannot be assumed. 

Second, we assume that only the total intensity of the radiation has an influence on the (polarized) populations of the molecular or atomic states. This is a reasonable assumption if the polarization fraction is low, which is corroborated by polarization observations of molecular emission lines. We refer to this approximation as the anisotropic intensity approximation. We discuss the validity of the anisotropic intensity approximation in more detail in Section \ref{sec:anis_int}, where we also compare our modeling with that of \citet{goldreich:81}, who take the influence of both the Stokes-I and -Q parameters on the alignment of the molecular states into account. 
 
These assumptions lead to significant simplifications in the theory behind the alignment of molecular and atomic quantum states and the radiation with which they interact. They allow for the implementation of such a model as an extension to a regular line radiative transfer code. In the following, we introduce the formalism that we used to model the alignment to molecular or atomic quantum states. After this, we outline how aligned quantum states influence the propagation of polarized radiation. 
\subsection{Polarized statistical equilibrium equations}
The polarizing mechanism we focus on is the anisotropic radiation field. Mathematically, anisotropy in the radiation field that affects the quantum state alignment is most easily described in terms of an irreducible tensor-element expansion. The irreducible tensor components of the radiation field, which are in direction $\Omega$ and at the position $\boldsymbol{r}$, are obtained as \citep{landi:84}
\begin{align}
\mathcal{J}^K_Q (\boldsymbol{r},\nu , \Omega) = \sum_j \mathcal{T}^K_Q(j,\Omega) S_j(\boldsymbol{r},\nu,\Omega),
\label{eq:irred_J}
\end{align}
where $K$ represents the irreducible tensor rank and $Q$ is its projection, $S_j (\nu,\Omega)$ are the Stokes-parameters at frequency $\nu,$ and $j$ runs over all four Stokes parameters. We define the Stokes parameters in relation to the complex electric field vector components as 
\begin{subequations}
\begin{align}
I = |E_x|^2 + |E_y|^2 , \\
Q = |E_x|^2 - |E_y|^2 , \\ 
U = 2\mathrm{Re}\left[E_x E_y^* \right], \\
V = 2\mathrm{Im}\left[E_x E_y^* \right], 
\end{align} 
\end{subequations}
where $x$ and $y$ refer to the axes that are perpendicular to the propagation direction, $z$, and each other. In this work, we consistently chose the axis of $x$ along the rejection of the (local) magnetic field direction from the propagation direction. The transformation coefficients $\mathcal{T}^K_Q(j,\Omega)$ are defined in equation~(A6) from \citet{landi:84}. If we only consider alignment by Stokes-I radiation and if we furthermore assume a dominant magnetic field, only the $K=0,2$ and $Q=0$ components are of interest \citep{landi:06}. Under these conditions, the irreducible tensor components of the radiation field reduce to
\begin{subequations}
\begin{align}
\mathcal{J}^0_0(\boldsymbol{r},\nu,\Omega) &= I(\boldsymbol{r},\nu,\Omega), \\
\mathcal{J}^2_0(\boldsymbol{r},\nu,\Omega) &= \sqrt{\frac{1}{2}}P_2 (\mu ) I(\boldsymbol{r},\nu,\Omega),
\end{align}
\label{eq:J_int}
\end{subequations}
where $\Omega = (\theta,\phi)$ is expressed in terms of the inclination and azimuth angles that are gauged with respect to the magnetic field direction. The quantity $P_2 (\mu)$ is the second-order Legendre polynomial and $\mu=\cos \theta$. The solid-angle integrated tensors at position $\boldsymbol{r}$ are readily obtained as
\begin{subequations}
\begin{align}
J^0_0 (\boldsymbol{r},\nu) &= \frac{1}{4\pi}\int_{-1}^1 d \mu \int_0^{2\pi} d\phi \  I(\boldsymbol{r},\nu,\mathrm{acos}(\mu),\phi), \\ 
J^2_0 (\boldsymbol{r},\nu) &= \frac{1}{4\pi \sqrt{2}} \int_{-1}^1 d \mu \ P_2 (\mu) \int_0^{2\pi} d\phi \  I(\boldsymbol{r},\nu,\mathrm{acos}(\mu),\phi). 
\end{align}
\label{eq:int_rad_tens} 
\end{subequations}
In the following, we refer to the ratio $J_0^0 (\boldsymbol{r},\nu) / J_0^2 (\boldsymbol{r},\nu)$ as the relative alignment of the radiation field. For an isotropic radiation field ($I(\boldsymbol{r},\nu,\Omega) = I(\boldsymbol{r},\nu)$), it should be noted that only the (isotropic) $J^0_0 (\boldsymbol{r},\nu)$-term survives.

Just as for the radiation field, we represent the molecular or atomic quantum states as irreducible tensor elements in order to most optimally utilize their symmetry properties. Quantum states are denoted as $\rho^K_Q (\alpha J)$, where $K$ is the rank of the irreducible tensor element and $Q$ is its projection. The total angular momentum of the associated quantum state is $J$ and all other quantum numbers characterizing the quantum state are collected in $\alpha$. The rank $K$ is positive and restricted to values of $\leq 2J$. The elements $K\geq 1$ of the population tensor relate to the alignment of the quantum state and the $K=0$ element relates to the population of the quantum state. Under the assumption of a strong magnetic field, we can neglect all but the $Q=0$ projection elements. Because of the symmetry of the radiation field, we only have to take elements into account where $K$ is even. \citet{landi:06} presented the statistical equilibrium equations for the polarized quantum state $\rho^K_0 (\alpha, J)$ under the following conditions: 
\begin{align}
\dot{\rho}^K_0 (\alpha J) &= \sum_{\alpha_l J_l K_l} \rho^{K_l}_0 (\alpha_l J_l) \left[ [t_A]_{\alpha J K}^{\alpha_l J_l K_l} + \sqrt{\frac{[J_l]}{[J]}} \delta_{K,K_l} [C_I^{(K)}]_{\alpha J}^{\alpha_l J_l} \right] \nonumber \\ 
&+  \sum_{\alpha_u J_u K_u} \rho^{K_u}_0 (\alpha_u J_u) \left[ [t_S]_{\alpha J K}^{\alpha_u J_u K_u} + [t_E]_{\alpha J K}^{\alpha_u J_u K_u} \right. \nonumber \\ 
&+ \left. \sqrt{\frac{[J_u]}{[J]}} \delta_{K,K_u} [C_S^{(K)}]_{\alpha J}^{\alpha_u J_u} \right] \nonumber \\
&-  \sum_{K'} \rho^{K'}_0 (\alpha J) \left[ [r_A]_{\alpha J K K'} + [r_E]_{\alpha J K K'} + [r_S]_{\alpha J K K'} \right. \nonumber \\
&+ \left. \delta_{KK'}\left( \sum_{\alpha_u J_u} [C_I^{(0)}]_{\alpha_u J_u}^{\alpha J} + \sum_{\alpha_l J_l} [C_S^{(0)}]_{\alpha_l J_l}^{\alpha J} + D^{(K)} (\alpha J)\right) \right].
\label{eq:stateq}
\end{align}
In Eq.~(\ref{eq:stateq}), the rate of radiative absorption events toward the $\rho^K_0 (\alpha, J)$ from lower level $ \rho^{K_l}_0 (\alpha_l J_l)$ is given by $[t_A]_{\alpha J K}^{\alpha_l J_l K_l}$ 
and the collisional contribution is $[C_I^{(K)}]_{\alpha J }^{\alpha_l J_l }$. The rate of stimulated and spontaneous emission events toward the $\rho^K_0 (\alpha, J)$ from upper level $ \rho^{K_u}_0 (\alpha_u J_u)$ are given by $[t_S]_{\alpha J K}^{\alpha_u J_u K_u}$ and $[t_E]_{\alpha J K}^{\alpha_u J_u K_u}$,
and the collisional contribution is $[C_S^{(K)}]_{\alpha J}^{\alpha_u J_u}$. The rates of absorption, stimulated emission, and spontaneous emission from the level $\rho^K_0 (\alpha, J)$ to all other levels is given by $[r_A]_{\alpha J K K'}$, $[r_S]_{\alpha J K K'}$, and $[r_E]_{\alpha J K K'}$.
Finally, the collisional depolarization rates are $D^{(K)}(\alpha J)$. More detailed expressions for the radiative rates from Eq.~(\ref{eq:stateq}) can be found in equations 7.20 from \cite{landi:06}. By assuming a steady-state, $\dot{\rho}^K_0 (\alpha J) = 0$, the statistical equilibrium equations can be solved as a linear set of equations. The solution yields the quantum state populations, including their relative alignment. 

We should note that the statistical equilibrium equations of Eq.~(\ref{eq:stateq}) are isomorphic to those presented in \citet{deguchi:84}. While \citet{deguchi:84} set up the statistical equilibrium equations in the standard angular momentum basis $\ket{jm}$, where $j$ is the total angular momentum of the eigenstate and $m$ is its projection, we worked in a spherical tensor representation. We refer to \citet{landi:06} for a detailed discussion on the relation between the two representations. We chose to work in a spherical tensor representation because of its symmetry properties. The properties of the spherical tensor expansion of both the molecular (or atomic) states and the radiation are such that truncation of higher-order $K$-terms in the $\rho^K_0 (\alpha J)$-expansion can be done with minimal loss of accuracy in the description of the statistical equilibrium equations for our system. Such truncation is not possible in the representation that \citet{deguchi:84} used, and it results in a rapid and unmitigable increase in computational effort when high angular momentum states are considered. 
\subsection{Polarized radiative transfer}
After having obtained the (aligned) quantum state populations, we can evaluate their impact on the radiation propagation. Because of the strong magnetic field, (locally) only Stokes-Q radiation is produced. 
The propagation of radiation around frequency, $\nu_{\alpha' J' , \alpha J}$, associated with a transition $\alpha' J' \to \alpha J$, can be described by
\begin{align}
\frac{d}{ds} \boldsymbol{I}_{\nu} = -\boldsymbol{\kappa}_{\nu}^{\alpha' J' , \alpha J} \boldsymbol{I}_{\nu} + \boldsymbol{\epsilon}^{\alpha' J' , \alpha J},
\label{eq:polrad} 
\end{align}  
where $\boldsymbol{I}_{\nu}=[I_{\nu},Q_{\nu},U_{\nu},V_{\nu}]$ is the Stokes vector and the propagation matrix 
\begin{align}
\boldsymbol{\kappa}_{\nu}^{\alpha' J' , \alpha J} = \begin{bmatrix}\eta_I^{\alpha' J' , \alpha J} (\nu) & \eta_Q^{\alpha' J' , \alpha J} (\nu) & 0 & \eta_V^{\alpha' J' , \alpha J} (\nu) \\ \eta_Q^{\alpha' J' , \alpha J} (\nu) & \eta_I^{\alpha' J' , \alpha J} (\nu) & 0 & 0 \\ 0 & 0 & \eta_I^{\alpha' J' , \alpha J} (\nu) & 0 \\ \eta_V^{\alpha' J' , \alpha J} (\nu) & 0 & 0 & \eta_I^{\alpha' J' , \alpha J} (\nu) \end{bmatrix} 
\label{eq:kappa_mat}
\end{align}
is significantly simplified if one assumes a dominant magnetic field. Because we only consider diamagnetic molecules with Zeeman splitting that are far weaker than the thermal broadening, the production of Stokes-V radiation through the Zeeman effect is negligible and we set $\eta_V^{\alpha' J' , \alpha J} (\nu) \to 0$. Thus, in PORTAL, we only consider the propagation of linearly polarized radiation. The expressions for the $\eta$-elements of Eq.~(\ref{eq:kappa_mat}) are \citep{landi:84} 
\begin{subequations}
\begin{align}
\eta_I^{\alpha' J' , \alpha J} (\nu) &= \frac{h \nu_{\alpha' J',\alpha J}}{4\pi} B_{\alpha' J' , \alpha J} \phi_{\nu_{\alpha' J' , \alpha J}}(\nu) \left\{  \left( \mathcal{N}_{\alpha' J'}  - \mathcal{N}_{\alpha J} \frac{[J']}{[J]}\right) \right. \nonumber \\
&+ \left. \left(\mathcal{N}_{\alpha' J'} w_{J'J}^{(2)} \sigma^2_0 (\alpha' J') - \frac{[J']}{[J]} \mathcal{N}_{\alpha J} w_{JJ'}^{(2)} \sigma^2_0 (\alpha J) \right) \right. \nonumber \\
& \times \left. \frac{3 \cos^2 \theta - 1}{2\sqrt{2}} \right\}, \\
\eta_Q^{\alpha' J' , \alpha J} (\nu) &= -\frac{h \nu_{\alpha' J',\alpha J}}{4\pi} B_{\alpha' J' , \alpha J} \phi_{\nu_{\alpha' J' , \alpha J}} (\nu) \left( \mathcal{N}_{\alpha' J'} w_{J'J}^{(2)} \sigma^2_0 (\alpha' J') \right. \nonumber \\ 
&- \left. \frac{[J']}{[J]}\mathcal{N}_{\alpha J} w_{JJ'}^{(2)}  \sigma^2_0 (\alpha J) \right) \frac{3\sin^2 \theta }{2\sqrt{2}},
\end{align}
\label{eq:eta}
\end{subequations}
where $\mathcal{N}_{\alpha J} = \mathcal{N} [J]^{1/2} \rho_0^0 (\alpha J)$ is the number density of the quantum state $\alpha J$, and $\phi_{\nu_{\alpha' J' , \alpha J}}$ denotes the normalized line-profile centered at $\nu_{\alpha' J' , \alpha J}$ in frequency-space. The symbols 
\begin{align}
w_{J'J}^{(2)} = (-1)^{1+J+J'}\sqrt{3[J']}\begin{Bmatrix} 1 & 1 & 2 \\ J' & J' & J\end{Bmatrix} \nonumber
\end{align}
were introduced by \citet{landi:84}. The quantity between curly brackets is a Wigner-6j symbol \citep{biedenharn:81}. We use the short-hand notation
$
\sigma_0^2 (\alpha J) = \rho_0^2 (\alpha J) / \rho_0^0 (\alpha J)
$ 
for the relative alignment of the quantum state $\alpha J$. The spontaneous emission events in the polarized radiative transfer equations of Eq.~(\ref{eq:polrad}) are represented in the $\boldsymbol{\epsilon}$-vector. The spontaneous emission contribution to the Stokes-U is zero in the strong magnetic field limit we consider. The Zeeman effect for diamagnetic molecules is way smaller than the thermal broadening, so we can set $\epsilon_V^{\alpha' J', \alpha J} \to 0$. The contributions to the Stokes-I and -Q parameters are \citep{landi:84}
\begin{subequations}
\begin{align}
\epsilon_I^{\alpha' J' , \alpha J} (\nu) &= \frac{h \nu_{\alpha' J',\alpha J}}{4\pi} A_{\alpha' J' , \alpha J} \phi_{\nu_{\alpha' J' , \alpha J}} (\nu) \nonumber \\
& \times \mathcal{N}_{\alpha' J'} \left\{  1  + w_{J'J}^{(2)} \sigma^2_0 (\alpha' J') \frac{3 \cos^2 \theta - 1}{2\sqrt{2}} \right\}, \\
\epsilon_Q^{\alpha' J' , \alpha J} (\nu) &= -\frac{h \nu_{\alpha' J',\alpha J}}{4\pi} A_{\alpha' J' , \alpha J} \phi_{\nu_{\alpha' J' , \alpha J}} (\nu) \nonumber \\
&\times\mathcal{N}_{\alpha' J'} w_{J'J}^{(2)} \sigma^2_0 (\alpha' J') \frac{3\sin^2 \theta }{2\sqrt{2}}.
\end{align}
\label{eq:eps}
\end{subequations}
\section{Methods}
 The formalism that we present in the previous section can be used to simulate the emergence of polarization in spectral lines using an (isotropic) excitation model as input. It can be directly used by assuming LTE excitation, or alternatively, the atomic and molecular excitation from any (3D) radiative transfer code can be input. We outline in the following how we used the LIME radiative transfer code \citep{brinch:10}. 

LIME is a Monte Carlo 3D radiative transfer code that works with a (weighted) randomly chosen grid. A physical structure can be input, whereupon a random grid is chosen that is weighed over the molecular density and other parameters \citep{ritzerveld:06}. After a number of Monte Carlo radiative transfer iterations, which are sped up by an accelerated lambda iteration \citep{rybicki:91}, the simulation converges on a molecular and atomic excitation over all of the nodes in the simulation. Subsequently, this solution can be ray traced to simulate an image of the physical structure under investigation. 

Rather than directly ray tracing the excitation solution, we used it to thoroughly map out the local anisotropy of the radiation field throughout the simulation. With the local anisotropy parameters of the radiation field, we modeled the polarized excitation of the molecular or atomic states under investigation. Having the polarized excitation mapped out throughout the simulation, we performed a polarized ray-tracing to obtain a polarized image of the physical structure under investigation. 

In the following, we outline in more detail how we implemented PORTAL. In the first paragraph, we discuss setting up the polarized statistical equilibrium equations using the output of a line radiative transfer code. In order to do this, we dedicated most of our attention to the mapping of the local anisotropy of the radiation fields. In the second paragraph, we detail the polarized radiative transfer that was performed in the polarized ray-tracing. Especially for simulations with nonuniform magnetic fields, it is crucial to pay extra attention to the frame of reference of the polarized radiation and the proper way to relate different frames of reference.        

 In PORTAL, we used the anisotropic intensity approximation and formulated the polarized statistical equilibrium equations in terms of irreducible tensor elements. This approach differs from other efforts such as LinePol \citep[][(submitted)]{kuiper:20}, which builds on LIME, is optimized for CO, and uses the formalism of \citet{goldreich:82} to describe the propagation of polarized radiation and its interaction with the molecular medium. LinePol takes two polarization modes of the radiation into account and uses a polarized accelerated lambda iteration scheme to obtain the state-populations in the simulation. At minimal cost to the accuracy of our results (see Sections \ref{sec:polsee} and \ref{sec:anis_int}), the approximations in PORTAL speed up the simulation tremendously and lead to the possibility to treat more complex systems. PORTAL allows for complex geometries, magnetic field structures, and the treatment of molecules with extensive energy structures.

\subsection{Polarized statistical equilibrium equations}
\label{sec:polsee}
The quantum state alignment is dependent on the local anisotropy of the radiation field, so it is important to obtain a good angular sampling of the radiation field at the location of the simulation nodes. Different angular integrations of Eqs.~(\ref{eq:int_rad_tens}) for the case of an internal source of radiation (e.g., a central stellar object) and the case of no internal radiation source were used. For the latter, the local angular integration was performed as
\begin{subequations}
\begin{align}
J^0_0 (\boldsymbol{r},\nu) &= \frac{1}{4\pi}\sum_{i=1}^{N_{\mu}} w_i^{\mu} \sum_{j=1}^{N_\phi(\mu_i)} w_j^{\phi} I(\boldsymbol{r},\nu,\mathrm{acos}( \mu_i),\phi_j),  \\ 
J^2_0 (\boldsymbol{r},\nu) &= \frac{1}{4\pi\sqrt{2}}  \sum_{i=1}^{N_{\mu}} w_i^{\mu} P_2 (\mu_i)  \sum_{j=1}^{N_\phi (\mu_i)} w_j^{\phi} I(\boldsymbol{r},\nu,\mathrm{acos}(\mu_i),\phi_j) 
\end{align}
\end{subequations}
where $\mu_i$ and $w_i^{\mu}$ are the coordinates and the weights, which were taken from the $N_{\mu}$-point Gaussian quadrature rule. The integration over $\phi$ was performed over $N_{\phi} (\mu)\propto \sqrt{1-\mu^2}$ equidistant points all with weight $2\pi/N_\phi$. 

In the case of an internal radiation source, it should be appreciated that the solid angle associated with the radiation coming from this internal source is well-defined. Therefore, the solid angle integration was divided up into rays coming from the internal radiation source; the number of rays is proportional to the solid-angle of the internal source $\Delta \Omega_* = \pi(|\boldsymbol{r}|/R_*)^2$ and all of the other rays were distributed equally over the remaining sphere surface.

The local radiation field parameters of a node at the position $\boldsymbol{r}$, summarized in $J^0_0 (\boldsymbol{r},\nu)$ and $J^2_0 (\boldsymbol{r},\nu)$, were obtained by ray tracing $N$ rays with direction $\boldsymbol{k}_{\mu,\phi}$ to that node. The parameters $\mu$ were chosen with respect to the magnetic field direction ($\boldsymbol{b} \cdot \boldsymbol{k}_{\mu,\phi} = \mu$). The angles $\phi$ were gauged with respect to a canonical direction not parallel to the magnetic field. The choice of the canonical direction is free as the angle $\phi$ is integrated out without weighing (see Eq.~\ref{eq:J_int}). The ray-tracing was performed using the molecular populations that were output by LIME, while also using some of the relevant LIME-input parameters, such as (local) temperature, (local) velocity, and gridding. The ray-tracing yielded the local radiation field parameters that were subsequently used to obtain the quantum state populations and alignment.

The quantum state populations and alignment were obtained from the statistical equilibrium equations (SEE) given in Eq.~(\ref{eq:stateq}). The SEE are a balance of the radiative and collisional transition events. The radiative transition events are dependent on local parameters for the (an)isotopic radiation field at frequencies of all of the allowed transitions and their associated Einstein coefficients. Collisional rates are dependent on the temperature-dependent collisional cross-sections and (local) number densities of the relevant collisional partners. 
%
%
The relevant Wigner coupling symbols were calculated using the WIGXJPF package \citep{johansson:16}. The SEE were formulated in terms of a set of linearly dependent equations and were subsequently solved via an LQ decomposition (using the LAPACK libraries, \citet{LAPACK}) under the following physical constraint: $\sum_i [j_i]^{1/2} \rho_0^0 (\alpha_i j_i) = 1$. The solutions also included the isotropic populations that were compared to the LIME-output. We found that neglecting the quantum state alignment terms, $\rho_0^k (\alpha j)$ with $k>2$, introduces an error of $\sim 1\% $ in the state-alignment expressions, and for $k>4$ this error is already reduced to $\sim 1\permil$. In general, the quantum state alignment can be neglected for terms, $\rho_0^k (\alpha j)$ with $k>6$, with virtually no loss in precision and with great reduction of computational effort as a consequence\footnote{For example, the dimensionality of the polarized SEE for the first 41 rotational levels of CO reduced from 861 to 151 by setting $k_{\mathrm{max}}=6$.}.

\subsection{Polarized radiative transfer}
The quantum state populations and alignment obtained from the SEE were used to compute the (polarized) absorption and emission factors for each node in the simulation. The angle $\theta$ in Eqs.~(\ref{eq:eta}-\ref{eq:eps}) was obtained from the local magnetic field direction and the ray-trace direction. The ray-trace direction was chosen by defining an inclination angle and azimuth angle. The polarized radiation was gauged with respect to a canonical axis, $\boldsymbol{\chi}_{\mathrm{global}}$, perpendicular to the ray-tracing direction. The local and global Stokes parameters are related as \citep{landi:06} 
\begin{align}
\begin{pmatrix} Q_{\mathrm{local}} \\ U_{\mathrm{local}} \end{pmatrix} = \begin{pmatrix} \cos 2\alpha \ & \sin 2\alpha \\ -\sin 2\alpha & \cos 2\alpha \end{pmatrix} \begin{pmatrix} Q_{\mathrm{global}} \\ U_{\mathrm{global}} \end{pmatrix},
\label{eq:globtoloc}
\end{align}
and $I_{\mathrm{local}} = I_{\mathrm{global}}$. In Eq.~(\ref{eq:globtoloc}), $\alpha$ is the angle between $\boldsymbol{\chi}_{\mathrm{global}}$ and $\boldsymbol{\chi}_{\mathrm{local}}$ and the local reference axis is the unit vector along the rejection of the local magnetic field direction from the ray-tracing direction. 

The local Stokes-parameters were propagated using the polarized radiative transfer equations. Equations~(\ref{eq:polrad}-\ref{eq:kappa_mat}) show that only the Stokes-Q and -I coefficients are coupled in the polarized radiative transfer. That means that the propagation of the Stokes-U radiation is simply $U(s) = U(0) e^{-\eta_I s}$. To evaluate the propagation of the other Stokes parameters, $\boldsymbol{i} = [I,Q]$, the evolution operator formalism of \citet{landi:06} was used, where the propagation is described by
\begin{align}
\boldsymbol{i}(s) = \int_0^s ds' \ \boldsymbol{O}(s,s') \boldsymbol{\epsilon}(s') + \boldsymbol{O}(s,0) \boldsymbol{i}(0),
\end{align} 
and where
\[
\boldsymbol{O}(s,s') = e^{\int_{s'}^s ds'' \  \eta_I} \begin{bmatrix} \cosh \left( \int_{s'}^s ds'' \ \eta_Q \right) & -\sinh \left( \int_{s'}^s ds'' \ \eta_Q \right) \\ 
 -\sinh \left( \int_{s'}^s ds'' \ \eta_Q \right) & \cosh \left( \int_{s'}^s ds'' \  \eta_Q \right) \end{bmatrix}
\]
is the evolution operator (see Chapter 8 of \citet{landi:84}). The propagation for each crossed cell was considered, and within such a propagation, the coefficients $\eta_I$ and $\eta_Q$ as well as $\epsilon_I$ and $\epsilon_Q$ are constant. It is then straightforward to evaluate the integrals inside the evolution operator as well as the integral over the evolution operator: $\int_0^s ds'\ \boldsymbol{O}(s,s')$. Having done so, the propagation of the Stokes-I and -Q within a single cell is given by 
\begin{subequations}
\begin{align}
I(s) &= o_{I} \epsilon_I + o_Q \epsilon_Q + \left[ \cosh (\eta_Q s) I(0) - \sinh(\eta_Q s) Q(0) \right]e^{-\eta_I s},  \\
Q(s) &= o_{Q} \epsilon_I + o_I \epsilon_Q + \left[ \cosh (\eta_Q s) Q(0) - \sinh(\eta_Q s) I(0) \right]e^{-\eta_I s}, 
\end{align} 
\end{subequations}
where
\begin{align}
o_I &= \frac{\eta_I}{\eta_I^2 - \eta_Q^2} \left(1 - \left[\cosh(\eta_Q s) + \frac{\eta_Q}{\eta_I} \sinh(\eta_Q s)\right] e^{-\eta_I s}\right), \nonumber \\
o_Q &= -\frac{\eta_Q}{\eta_I^2 - \eta_Q^2} \left(1 - \left[\cosh(\eta_Q s) + \frac{\eta_I}{\eta_Q} \sinh(\eta_Q s)\right]e^{-\eta_I s} \right) \nonumber 
\end{align}
are the factors that were obtained from integrating the elements of the evolution operator.

\section{Simulations}
We applied PORTAL to known astrophysical problems. We consider the standard problem of a spherically symmetric collapsing molecular cloud, and we investigate the emergence of polarization in molecular lines through an anisotropic radiation field in a standard protoplanetary disk system. It should be noted that neither of these problems illustrate the full 3D capabilities of PORTAL. We focus, however, on these models because of their more straightforward interpretation and we leave more complex modeling for further work.
 
\subsection{Collapsing spherical cloud}
A benchmark problem in radiative transfer modeling, which furthermore allows for local anisotropy to establish itself in the radiation field, is the problem of a collapsing spherical cloud. We consider the emergence of polarization in HCO$^+$ lines. The density, velocity, and temperature distribution are taken from the \citet{shu:77} collapse model, using the same parameters as \citet{zadelhoff:02}. Only the ground vibrational state of HCO$^+$ is considered. We assume a uniform HCO$^+$ abundance of $10^{-9}$ and assume constant turbulent broadening of $200 \ \mathrm{m/s}$. We assume that a strong radial magnetic field (origin: center of mass) permeates the cloud. 

\begin{figure}[h!]
  \centering
  \begin{subfigure}[b]{0.45\textwidth}
    \includegraphics[width=\textwidth]{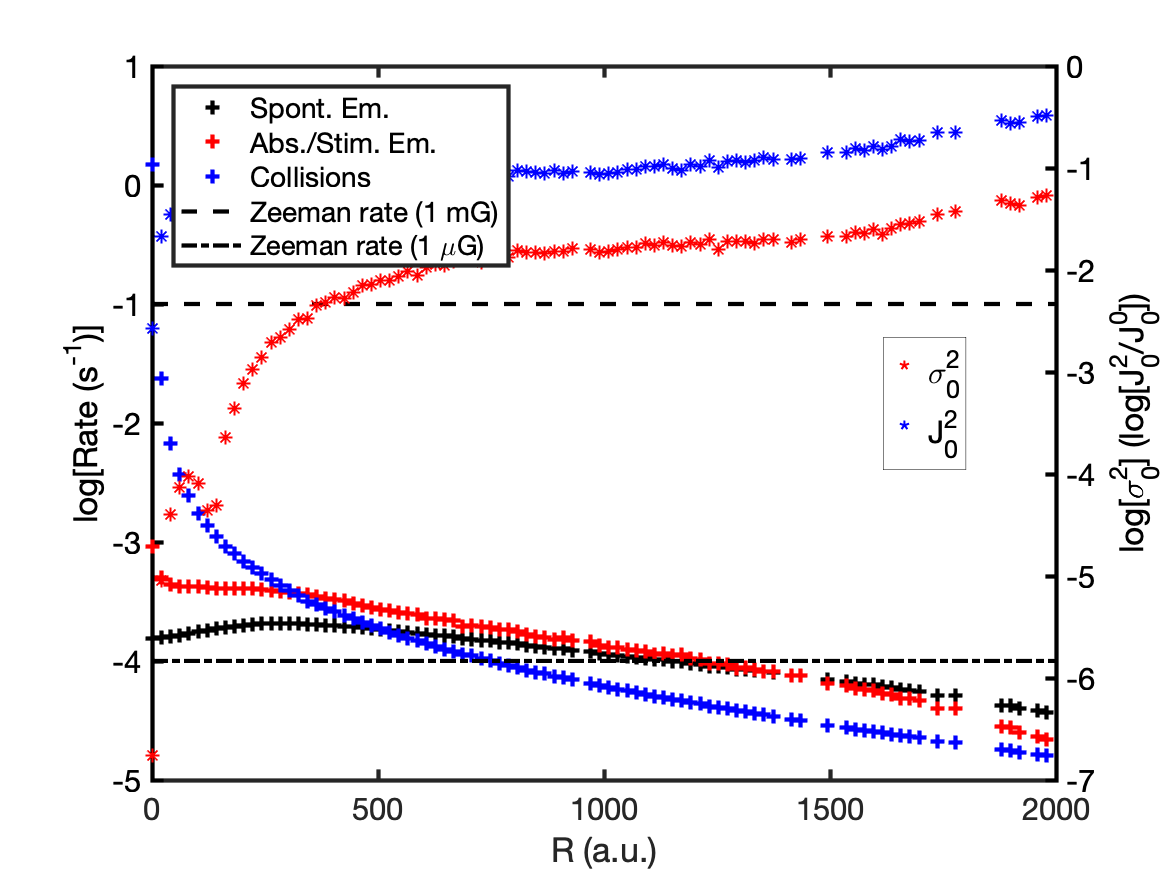}
    \caption{}
  \end{subfigure}
  \begin{subfigure}[b]{0.45\textwidth}
    \includegraphics[width=\textwidth]{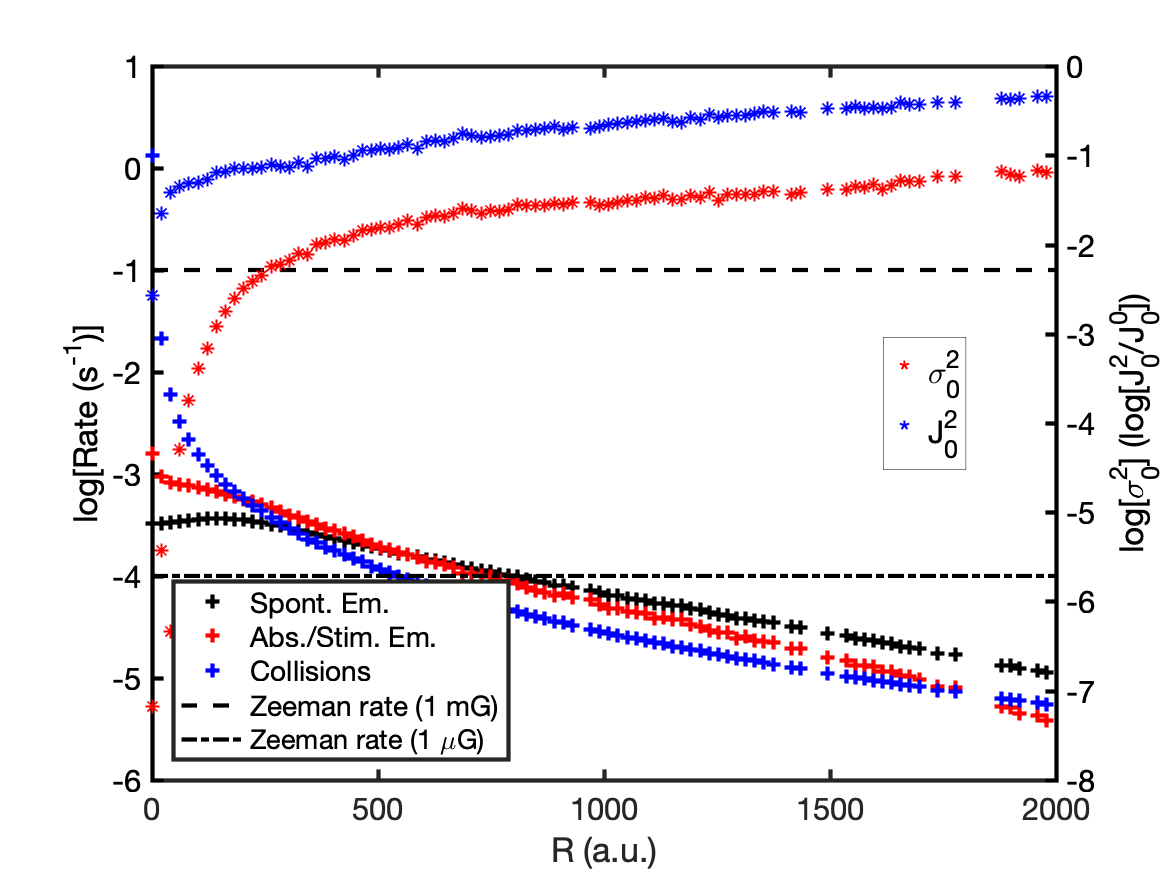}
    \caption{}
  \end{subfigure}
  \caption{Plot of the collapsing sphere's interaction rates (collisional, absorption, and stimulated emission as well as spontaneous emission) and relative alignment (radiative and quantum state) as a function of the radius for (a) the $J=2$ level and $J=2-1$ transition and (b) the $J=3$ level and $J=3-2$ transition. The interaction rates should be read from the left axis, the relative alignment from the right axis.}
  \label{fig:sphere_rates}
\end{figure}

First of all, an overview of the relevant isotropic and anisotropic interactions is instrumental to an eventual discussion of the quantum state alignment and radiation polarization characteristics. We report the cumulative radiative and collisional rates of the $J=2$ and $J=3$ level of HCO$^+$ in Figure \ref{fig:sphere_rates}. Of the different interactions, only stimulated emission and absorption are anisotropic interactions. Using the spherical symmetry of the collapsing sphere-problem, we only plotted the rates as a function of the distance to the center. We observe that for the inner regions of the collapsing sphere, collisions become dominant as the density of this regions increases. Even though there is appreciable alignment of the radiation field, the quantum states do not align themselves because of the dominant isotropizing collisions. From about 400 AU, radiative interactions take over as the dominant interaction and the quantum states align themselves. We also give the magnetic precession rate for a magnetic field of $1$ mG and 1 $\mathrm{\mu G}$ and note that for a HCO$^+$-molecule in the collapsing sphere, the magnetic field can be taken to define the symmetry axis when it is $\sim 10-100 \times$ stronger than other interactions. From Figure \ref{fig:sphere_rates}, we estimate this to be the case at magnetic field strengths of $\sim 10-100 \ \mathrm{\mu G}$.

In the same Figure \ref{fig:sphere_rates}, we plotted the relative anisotropy of the radiation field and the relative alignment of the quantum states $J=2$ and $J=3$. We note that the radiation anisotropy increases, thus moving away from the collapsing-sphere center. The radiation anisotropy in the collapsing sphere is partly a result of the density structure and partly the result of the velocity structure. Both structures are spherically symmetric, but this spherical symmetry is only manifest when the center is taken as the origin. For any cell that is not located at the center of the collapsing sphere, the radiation field is therefore anisotropic. Higher anisotropy in the radiation is associated with a stronger alignment of the quantum states.

\begin{figure}[h!]
  \centering
  \includegraphics[width=0.5\textwidth]{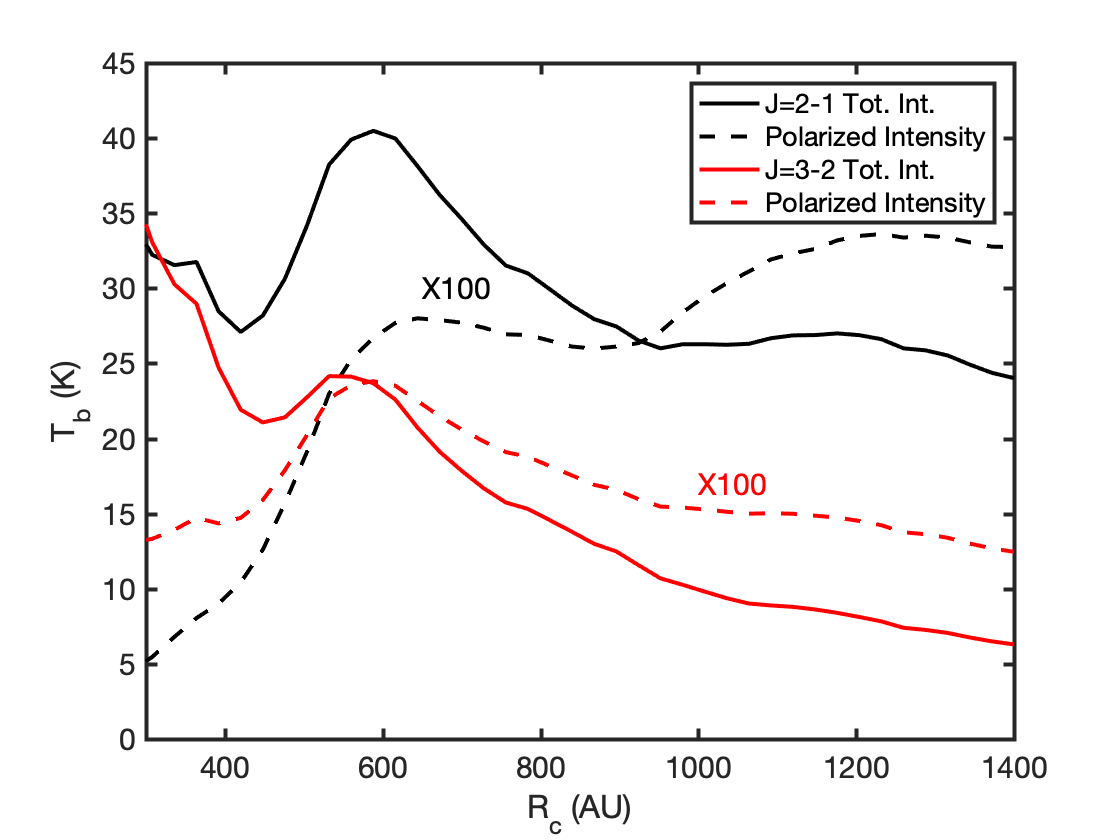}
  \caption{Total and polarized emission intensity (in Kelvin) of a collapsing sphere as a function of the radial distance.}
  \label{fig:sphere_frac}
\end{figure}

\begin{figure}[h!]
  \centering
  \begin{subfigure}[b]{0.45\textwidth}
    \includegraphics[width=\textwidth]{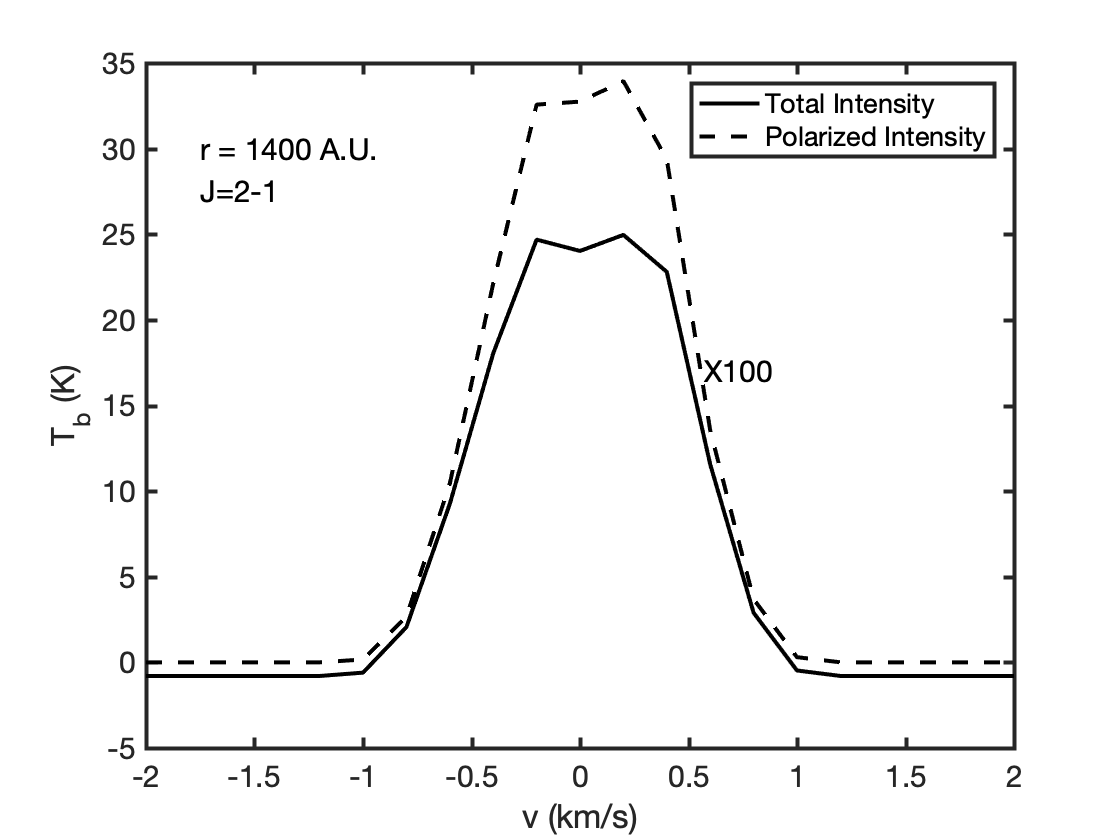}
    \caption{}
  \end{subfigure}
  \begin{subfigure}[b]{0.45\textwidth}
    \includegraphics[width=\textwidth]{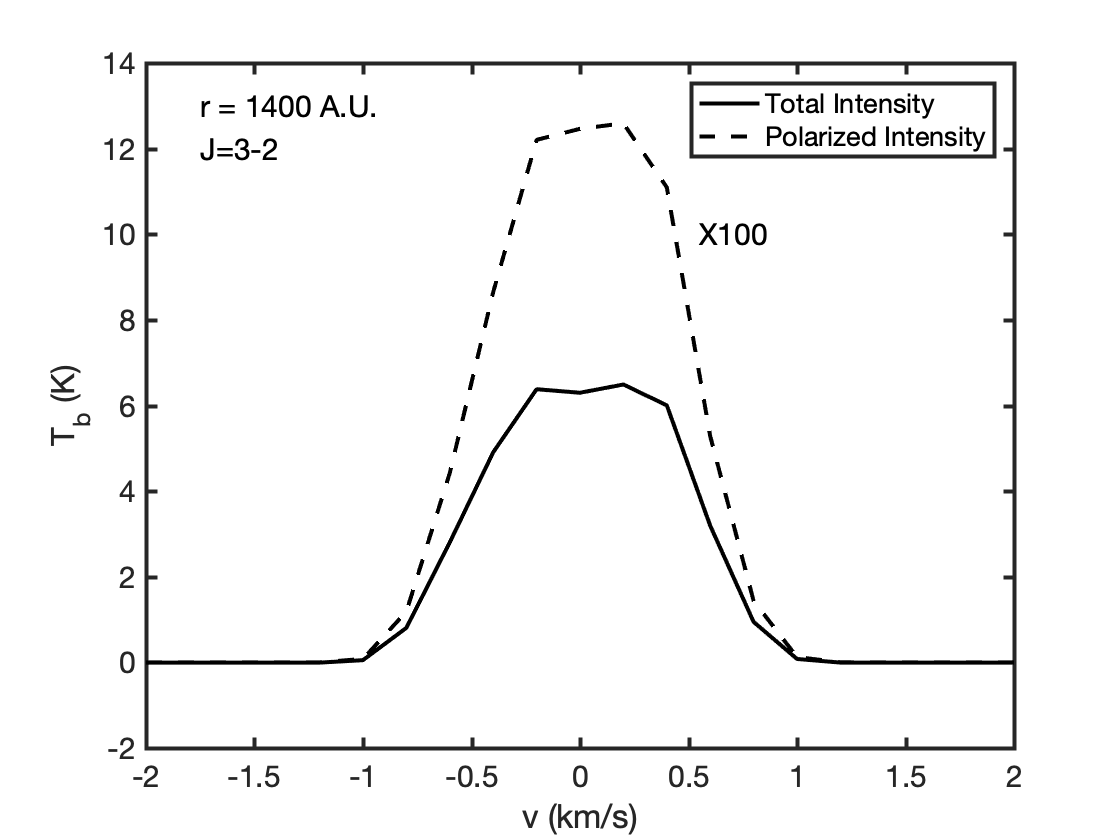}
    \caption{}
  \end{subfigure}
  \caption{Spectra of the (polarized) intensity (in Kelvin) of the (a) $J=2-1$ and (b) $J=3-2$ transitions from a collapsing sphere. The spectra are azimuthally averaged at $1400$ AU.}
  \label{fig:sphere_spec}
\end{figure}

We report the azimuthally averaged total intensity and polarization fraction of the HCO$^+$ $J=3-2$ and $J=2-1$ transitions in Figure \ref{fig:sphere_frac}. Indeed, we note that close to the center of the collapsing sphere, the polarization fraction is the lowest and gradually increases when moving outward. Polarization fractions are above $1 \%$ for a radial distance greater than $600$ AU for the $J=3-2$ transition and $900$ AU for the $J=2-1$ transition. We report the associated spectra at $R=1400$ AU in Figure \ref{fig:sphere_spec}. We observe that the linear polarization spectra roughly follow the spectral shape of the total intensity. 

The polarization angles are oriented in a radial fashion along the magnetic field lines. One should be aware that for a radial magnetic field, the angle between the magnetic field lines and the propagation direction toward the observer is a function of the propagation position. Accordingly, at the magic angle of $\theta_{\mathrm{magic}} \approx 54.7$ or $z/R = \frac{1}{\sqrt{3}}$, the propagation elements $\eta_Q$ flip sign and some of the earlier produced polarization is negated.
\subsection{Protoplanetary disk}
The protoplanetary disk is a prime example of an anisotropic astrophysical structure. Both the anisotropy in the density and velocity structure produce a locally anisotropic radiation field. Magnetic fields in the protoplanetary disk have been conjectured through dust-polarization observations \citep{stephens:17}, and recently, stringent limits have been put on the magnetic field strength through ALMA line circular polarization observations \citep{vlemmings:19}. 

We consider the polarization of $^{12}$CO in a general toy model of a protoplanetary disk having a number-density distribution of 
\begin{align}
n_{\mathrm{H}_2}(r_c,h) = 4\times 10^{14} \left(\frac{h}{\mathrm{AU}}\right)^{-2.25} e^{-50 \frac{(h/\mathrm{AU})^{2}}{(r_c/\mathrm{AU})^{2.5}}} \ \mathrm{m}^{-3}, 
\end{align}
where $r_c$ is the radial distance and $h$ is the height. The disk is assumed to be rotating, resulting in a model velocity-field of
\begin{align}
\boldsymbol{v}(\boldsymbol{r})= v(\cos \phi \hat{\boldsymbol{x}} - \sin \phi \hat{\boldsymbol{y}}), 
\end{align}
where
\[
v = 2.11 \times 10^4 \left(\frac{r_c}{\mathrm{AU}}\right)^{-1} \ \mathrm{m/s},
\]
and $\tan \phi = y/x$. The temperature is given by
\begin{align}
T(r_c) = 400 \left(\frac{r_c}{\mathrm{AU}} \right)^{-\frac{1}{2}} \ \mathrm{K}.
\end{align}  
Furthermore, we assume a constant CO abundance of $10^{-3}$ and a constant turbulent doppler broadening of $b_{\mathrm{turb}}=200 \ \mathrm{m/s}$. We only take the vibrational ground-state of $^{12}$CO into account. We neglect any line-overlap with transitions from other species. We explore the emergence of polarization in a protoplanetary disk for three types of (strong) magnetic fields: radial, toroidal, and poloidal.

We note that perhaps this toy model of the protoplanetary disk does not capture all features of the protoplanetary disk that are important in considering the polarization of thermal lines. For instance, we neglect to represent the inner midplane region by optically thick dust, so that the anisotropic radiation field resulting therefrom is not accounted for. Also, by not taking vibrationally excited levels and the transitions between different vibrational levels into account, we fail to include their significant aligning interactions (see Section \ref{sec:comp_align}). We explore more detailed and thorough modeling of protoplanetary disk regions in future work. These results should be taken as a simplified, but generally indicative, model of the mechanisms involved in the polarization of thermal line radiation of radiation by a magnetic field in protoplanetary disk regions. 
                                 
\begin{figure}[h!]
  \centering
  \begin{subfigure}[b]{0.45\textwidth}
    \includegraphics[width=\textwidth]{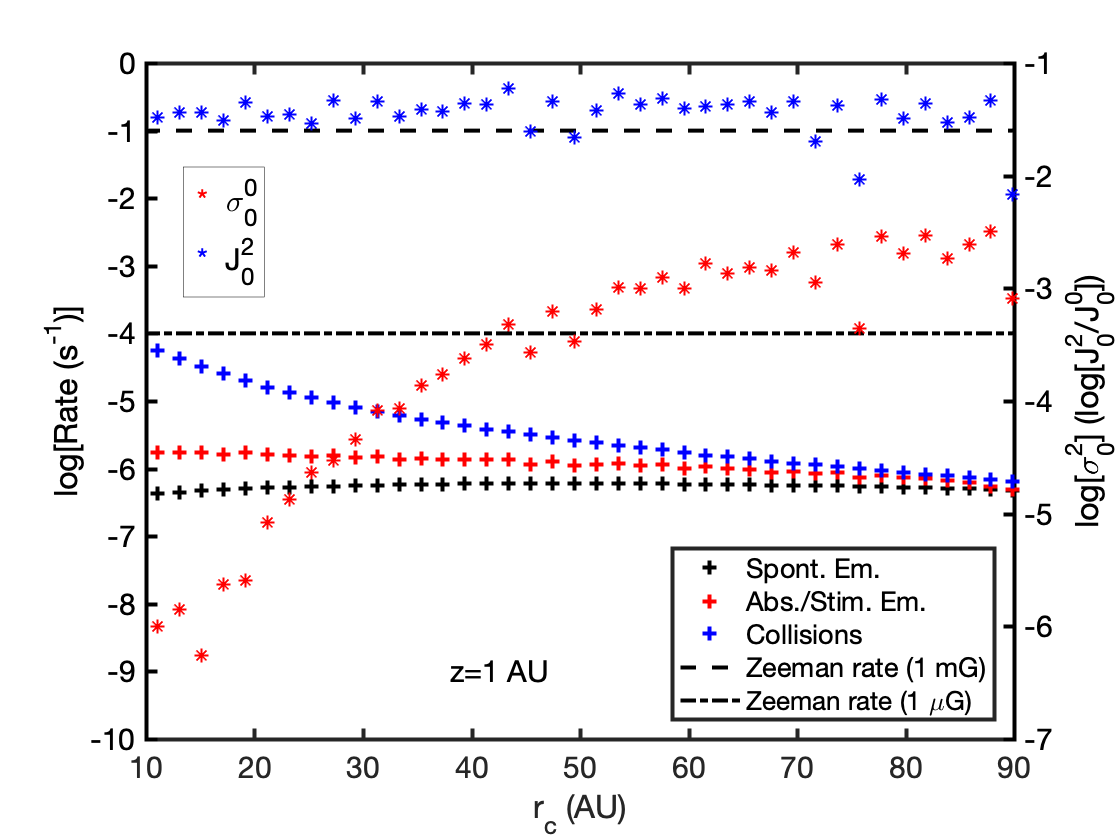}
    \caption{}
  \end{subfigure}
  \begin{subfigure}[b]{0.45\textwidth}
    \includegraphics[width=\textwidth]{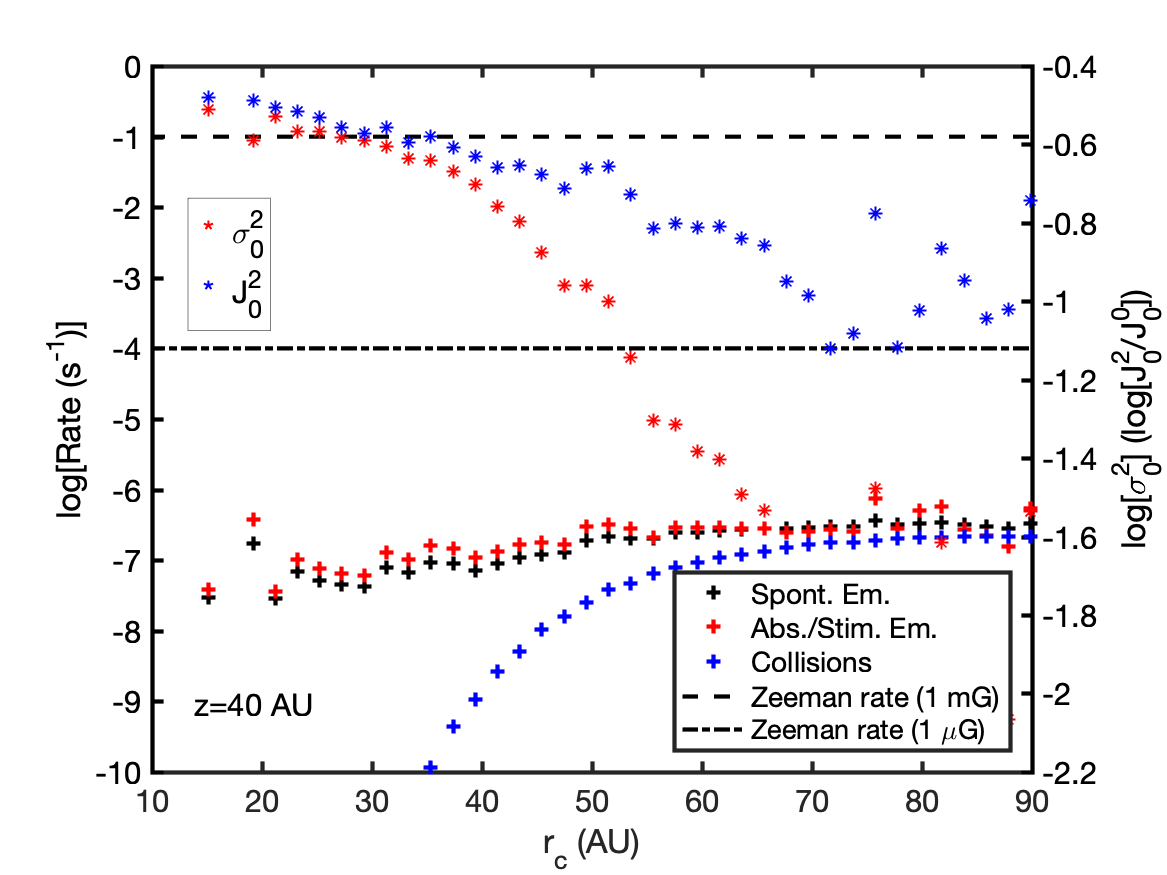}
    \caption{}
  \end{subfigure}
\caption{Plot of the protoplanetary disk's interaction rates (collisional, absorption, and stimulated emission as well as spontaneous emission) and relative alignment (radiative and quantum state, with respect to a toroidal magnetic field) as a function of the radial distance for (a) $1$ AU height and (b) $40$ AU height. The interaction rates should be read from the left axis, the relative alignment from the right axis.}
  \label{fig:pp_rates}
\end{figure}

It is important to map out the rates of isotropic and anisotropic interactions in order to understand the relative alignment of the molecules or atoms. Because of the cylindrical symmetry of the protoplanetary disk, we are able to analyze the interaction rates as a function of the radial distance and the height. In Figure \ref{fig:pp_rates}, we report the cumulative radiative and collisional rates for the $J=3$ level of CO. The rates are plotted as a function of $r_c$ for different height-cross sections. We also report the magnetic precession rate of a $1\ \mathrm{\mu G}$ and a $1 \ \mathrm{mG}$ magnetic field. It is apparent that magnetic interactions dominate other interactions and that we are justified in choosing the projection-axis along the magnetic field direction. Further, we observe a dominance of collisions over other interactions in a large region of the inner parts of the protoplanetary disk. In the disk midplane, isotropic collisions dominate the radiative interactions in the disk, but this dominance becomes weaker with the radial distance. In the outer parts of the disk, where the density drops, collisions become weaker and radiative events dominate.

In Figure \ref{fig:pp_rates} we also plotted the relative anisotropy of the radiation field resonant with the $J=3-2$-transition and the relative alignment of the $J=3$ state. Both of these parameters are defined with respect to a toroidal magnetic field configuration. The radiation anisotropy is strongest in the outer parts of the disk and weakest in the bulk of the disk. The same dependence is seen for the relative alignment of the quantum states. The relative anisotropy of the radiation is almost constant as a function of the radial distance at a height of $1$ AU. This is because the disk is optically thick in the midplane. The local angular radiation profile is not isotropic because of the temperature gradient. Due to dominant collisions, the quantum state alignment in the midplane is not large enough to significantly polarize radiation that is coming through. 

\begin{figure*}[h!]
  \centering
  \begin{subfigure}[b]{0.45\textwidth}
    \includegraphics[width=\textwidth]{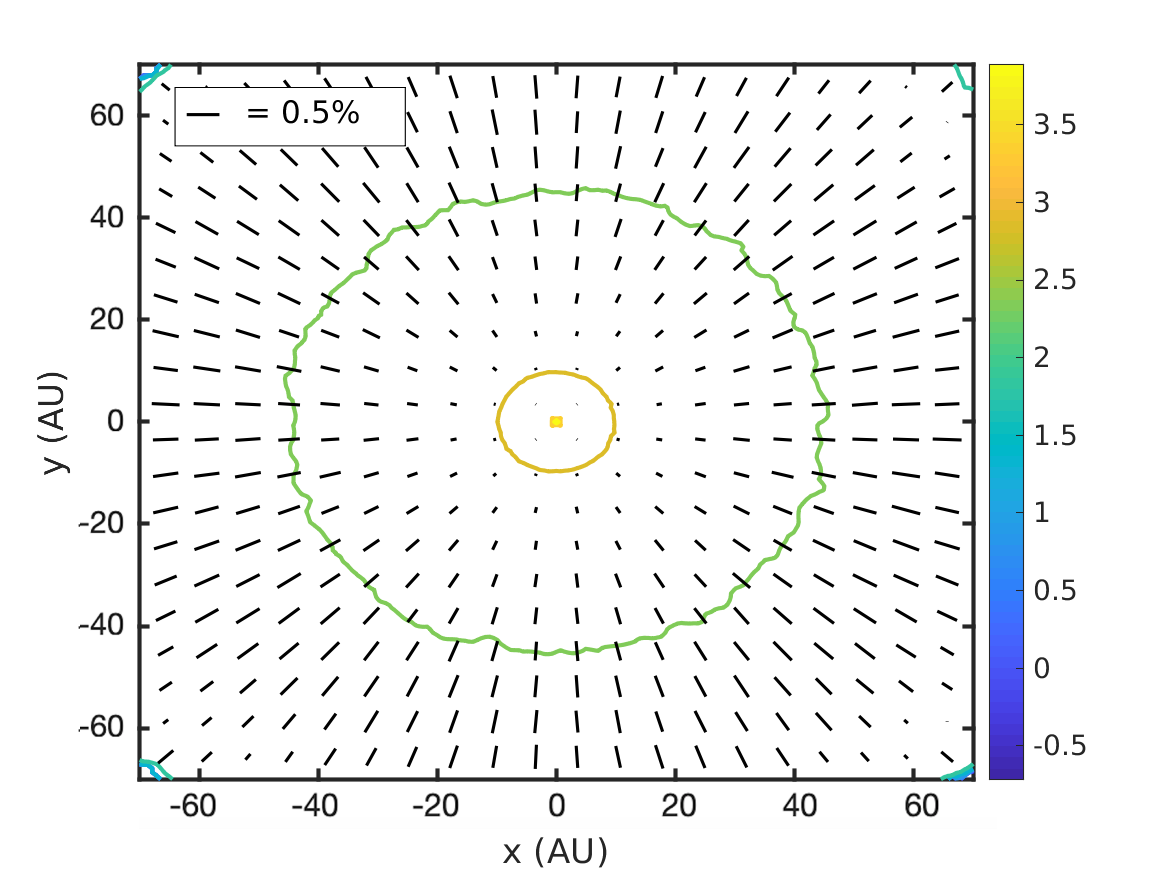}
    \caption{}
    \label{fig:pp_contour_rad}
  \end{subfigure}
  ~
  \begin{subfigure}[b]{0.45\textwidth}
    \includegraphics[width=\textwidth]{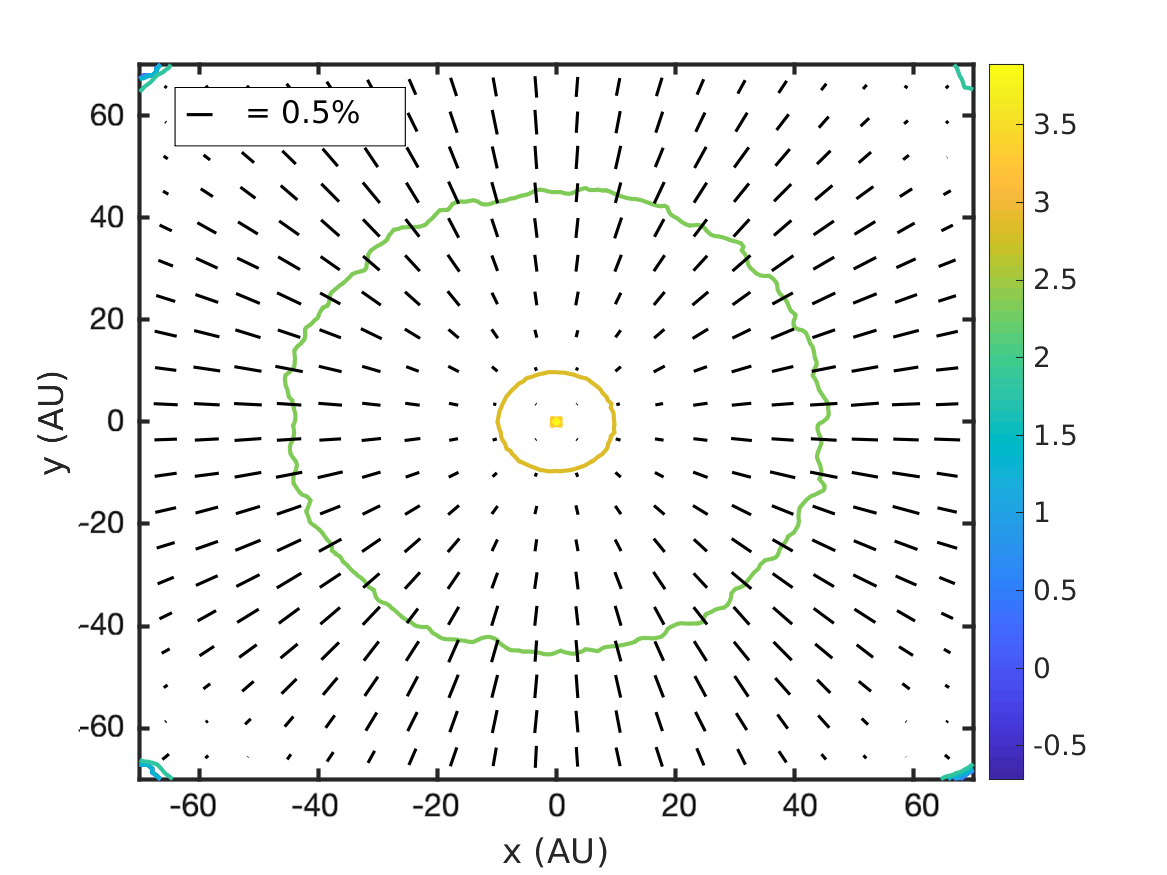}
    \caption{}
    \label{fig:pp_contour_tor}
  \end{subfigure}
  \begin{subfigure}[b]{0.45\textwidth}
    \includegraphics[width=\textwidth]{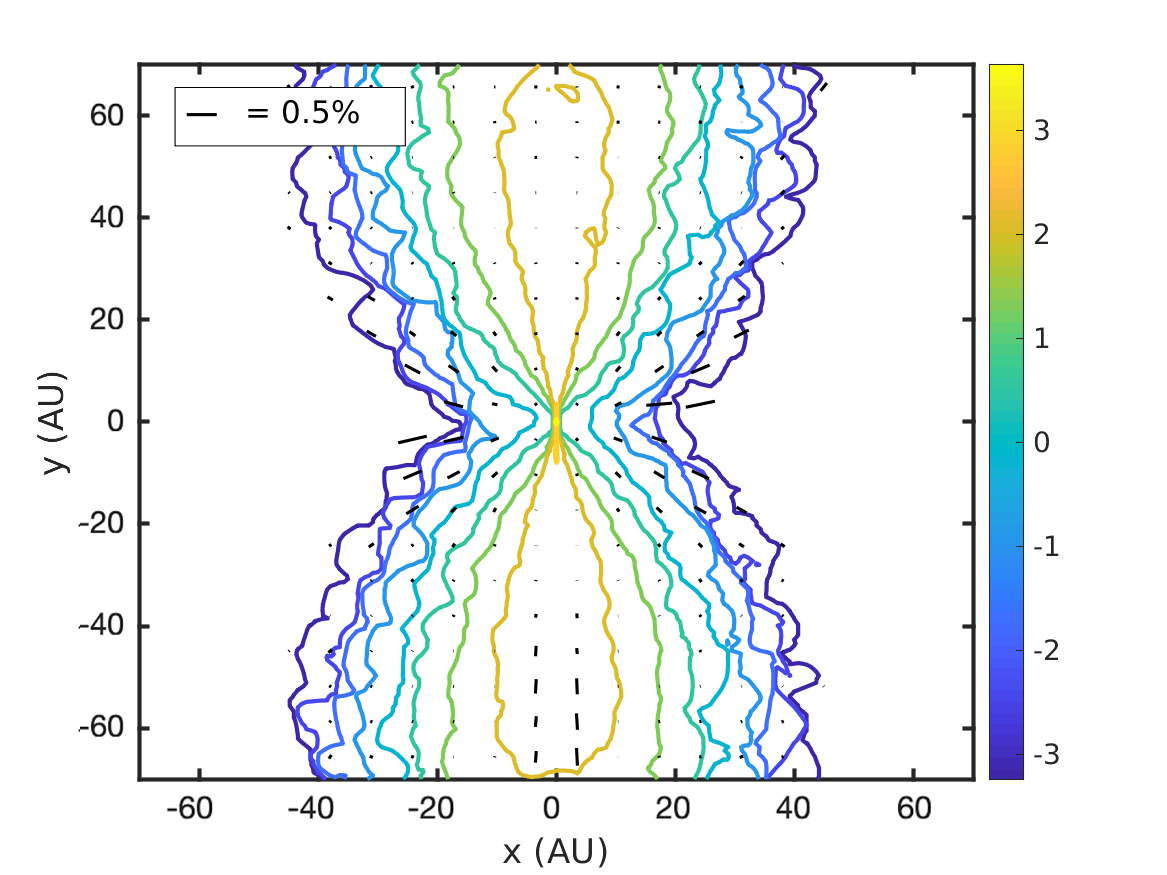}
    \caption{}
    \label{fig:pp_angle_rad}
  \end{subfigure}
  ~
  \begin{subfigure}[b]{0.45\textwidth}
    \includegraphics[width=\textwidth]{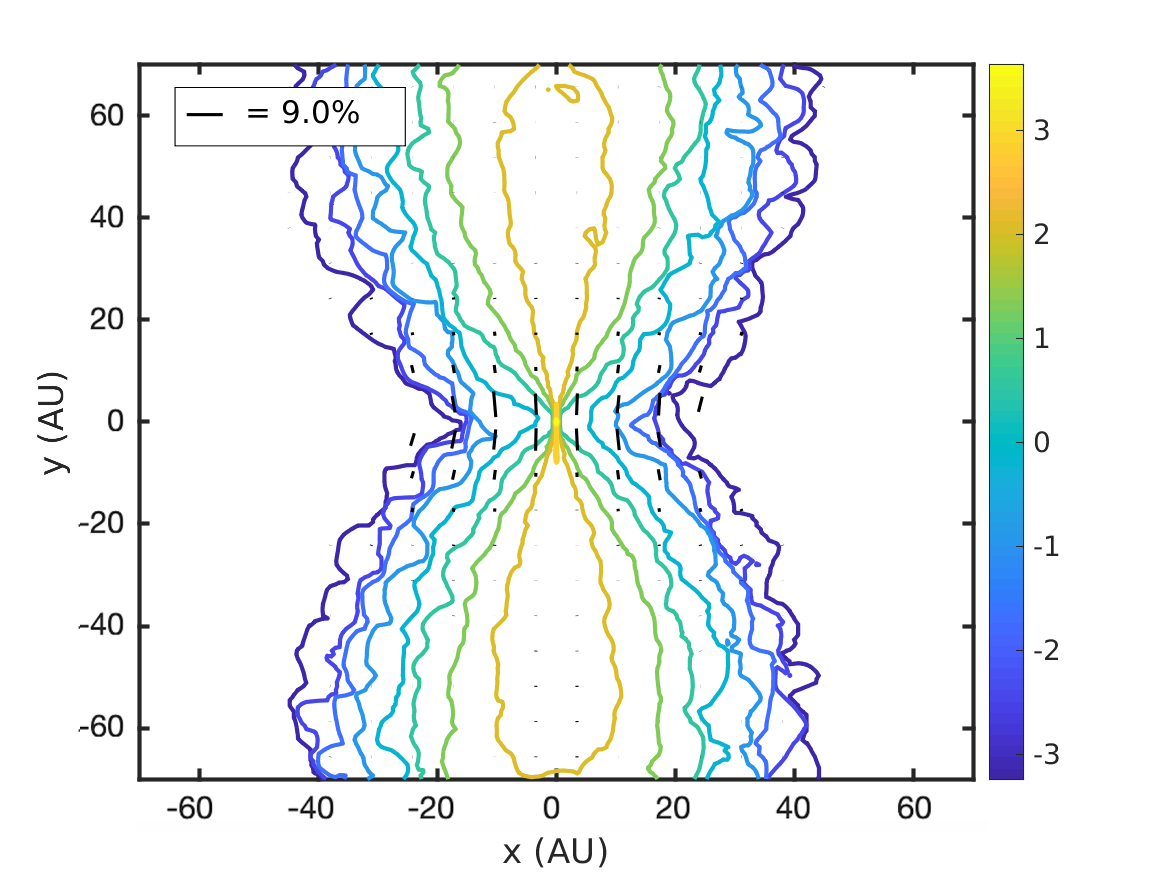}
    \caption{}
    \label{fig:pp_angle_tor}
  \end{subfigure}
  \caption{Contour plots of (the logarithm of) the total intensity (in Kelvin) of a protoplanetary disk. The disk is viewed face on [(a) and (b)] and at an inclination of $45^o$ [(c) and (d)]. We overlayed the intensity plot with polarization vectors from PORTAL simulations that come from a radial magnetic field (a,c) and a toroidal magnetic field (b,d). Polarization vector lengths scale with the polarization fraction.}
  \label{fig:pp_contour}
\end{figure*}

\begin{figure}[h!]
  \centering
  \includegraphics[width=0.5\textwidth]{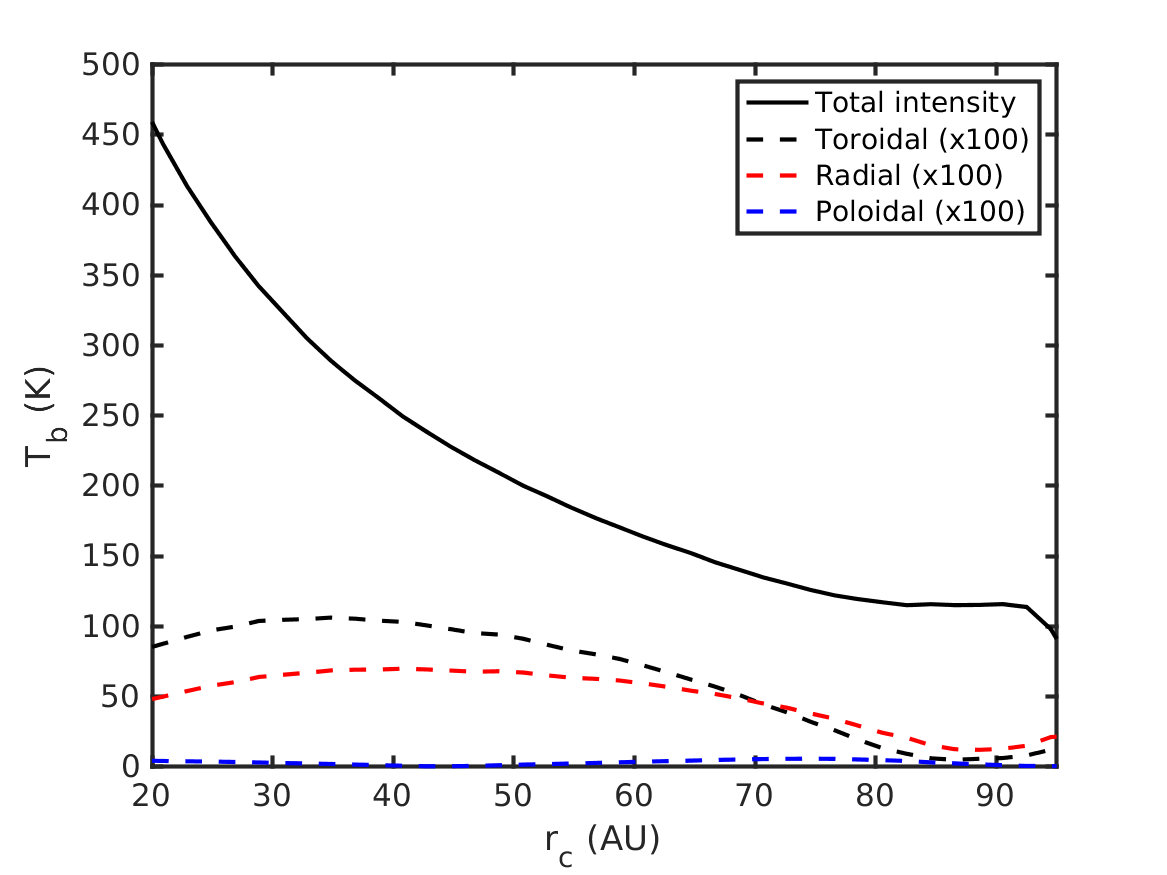}
  \caption{(Polarized) emission intensity (in Kelvin) of a protoplanetary disk as a function of the radial distance. The polarized intensity is plotted for three different magnetic field configurations and the disk is seen face on.}
  \label{fig:pp_frac}
\end{figure}

We analyzed the emergence of polarization through two different magnetic field configurations: toroidal and radial. In Figure \ref{fig:pp_contour_rad} we report the contour map of the $J=3-2$ CO-transition at $345.8$ GHz of the total intensity (in Kelvins) overlayed with polarization vectors resulting from the polarized emission of CO aligned with a radial magnetic field. The polarization vectors are scaled with respect to the polarization fraction and are parallel to the radial configuration of the magnetic field. Figure \ref{fig:pp_contour_tor} gives the polarization map coming from a toroidal magnetic field. We note that the polarization fraction for the face-on view of the protoplanetary disk is cylindrically symmetric. 

It is striking that the polarization vector maps viewed face on, for both the toroidal and radial magnetic field, have the same configurations. This similarity can be traced back to the anisotropy introduced in the molecular states via the anisotropic radiation field, $J^2_0 (\boldsymbol{r})$ (Eq.~\ref{eq:int_rad_tens}). When performing the integration to acquire $J^2_0 (\boldsymbol{r})$, the $\mu$-angle is gauged with respect to the magnetic field direction. The different gauges with respect to the toroidal and radial magnetic field configurations lead to the $J^2_0 (\boldsymbol{r})$, which is associated with the toroidal magnetic field, to be negative, while the $J^2_0(\boldsymbol{r})$ of the radial magnetic field is positive. Thus, in the region where polarization is produced, where furthermore the angle between propagation and the magnetic field $\theta_{\mathrm{prop}} > \theta_{\mathrm{magic}}$ for both magnetic field configurations, this gives rise to perpendicular and parallel orientations of the polarization vectors with respect to the toroidal and radial magnetic fields, that is,~polarization vectors that are identically oriented. Only when we view the disk at a significant inclination are we able to discern the orientation of the magnetic field from its polarization vectors, which can be seen in Figures \ref{fig:pp_angle_tor} and \ref{fig:pp_angle_rad}. 

The polarization maps of a protoplanetary disk viewed at a $45^o$ inclination show large polarization fractions for the poloidal and toroidal magnetic field configurations. Lower but still significant polarization fractions are seen to emerge from the radial magnetic field configuration. The highest polarization fractions occur at the edges of the protoplanetary disk. In the disk midplane, almost no polarization arises. This effect can be ascribed to the high optical depth from this region; it should, however, also be noted that our method underestimates the polarization fraction coming from optically thick regions (see Section \ref{sec:anis_int}). 

For the face-on view of a protoplanetary disk that is permeated by a poloidal magnetic field, no significant polarization emerges even though the quantum states are aligned. This is because for a large part of the disk, the magnetic field is almost aligned along the propagation direction. When this is the case, the propagation coefficients are $\eta_Q \to 0$, and no polarization is produced. When the disk is viewed at a significant inclination, the poloidal magnetic field produces a large polarization fraction. 

\begin{figure}[h!]
  \centering
  \includegraphics[width=0.5\textwidth]{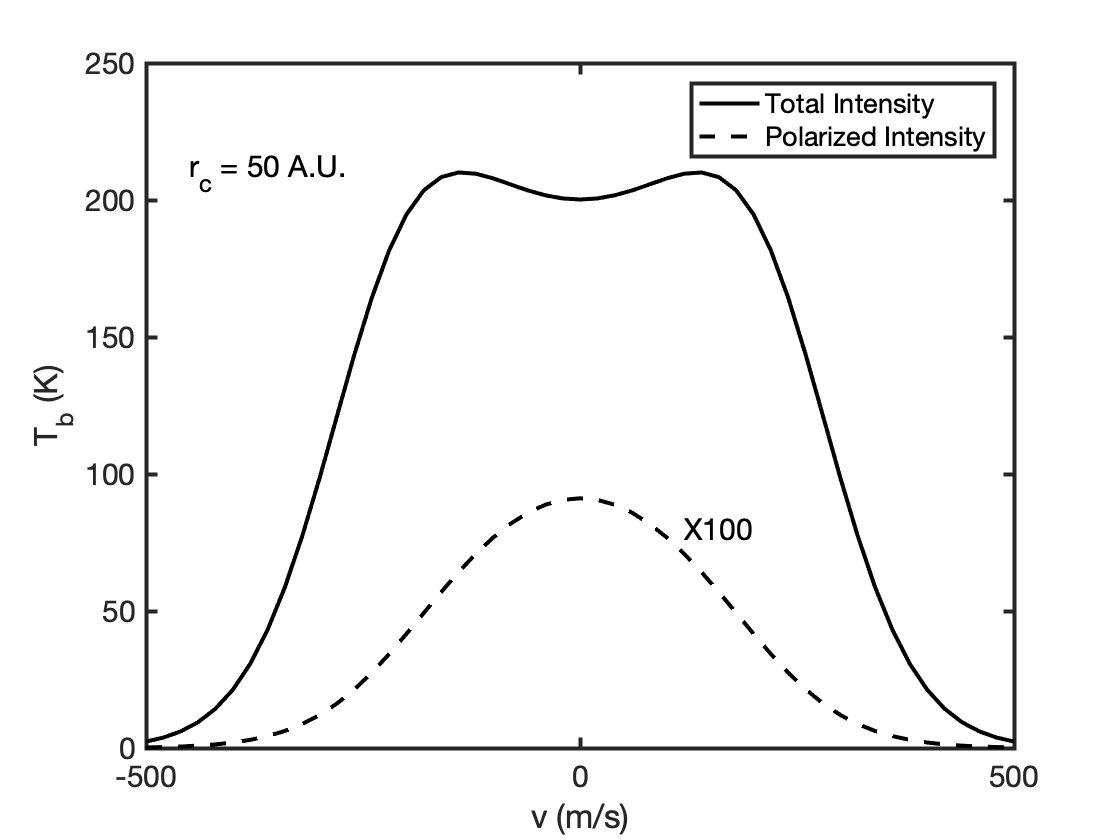}
  \caption{Spectrum of the (polarized) intensity (in Kelvin) of the $J=3-2$ transition from a protoplanetary disk permeated by a toroidal magnetic field. The spectrum is azimuthally averaged at $60$ AU and the disk is seen face on.}
  \label{fig:pp_spec}
\end{figure}

Figure \ref{fig:pp_frac} is a plot of the azimuthally averaged polarization fraction as a function of the radial distance. Near the center of the proto-planetary disk, the polarization fraction is low and increases as one moves outward. The maximum polarization fraction of the protoplanetary disk viewed face on is  $\sim 0.5 \%$, but polarization fractions up to $\sim 9\%$ are observed when the disk is viewed at an inclination of $45^o$. We analyze the azimuthally averaged ($r_c = 50$ AU) spectrum of the total (polarized) intensity in Figure \ref{fig:pp_spec}. The polarization roughly follows the spectral shape of the total intensity.

It is a general trend that high-frequency transitions have a larger tendency to emit polarized radiation. This is because the radiative rates scale with the frequency. Radiative interactions of high-frequency transitions therefore tend to dominate over collisional interactions. At the same time, the transition optical depth falls (generally) with the transition frequency; for transitions that are too optically thin, radiation intensity is too low to align the quantum states. 

\section{Discussion}
The anisotropic intensity approximation and the strong magnetic field approximation are central to the quality of the method we employed in PORTAL. We discuss these two approximations in the following two subsections. We discuss general remarks about the simulations of astrophysical regions using PORTAL in Section \ref{sec:general_remarks}. 
\subsection{The anisotropic intensity approximation}
\label{sec:anis_int}
\begin{figure}[h!]
  \centering
  \includegraphics[width=0.5\textwidth]{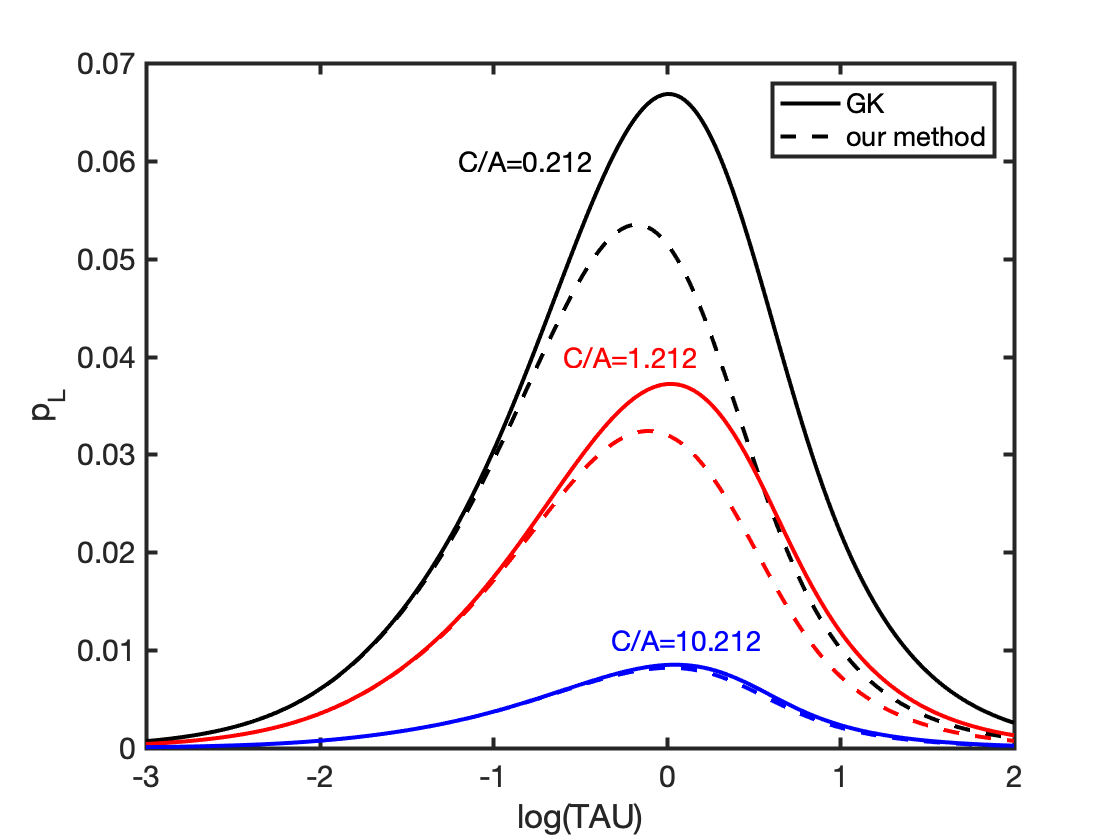}
  \caption{Comparison of the polarization fraction computed through the GK method (solid line) and the radiation anisotropy method we employ in this paper (dotted line). For more details on the simulation parameters, see \citet{goldreich:81}. We consider a $J=1-0$ transition at $100$ GHz, with a strong magnetic field along the $\hat{z}$-axis and a velocity gradient of $10^{-9}$ s$^{-1}$ in the $xy$-plane. We consider a temperature of $T=10$ K. Three ratios for the collision-radiative rates are considered and denoted inside the figure. The polarization fraction was computed for a ray traveling along the $\hat{x}$-axis.}
  \label{fig:GK_compare}
\end{figure}

Our method heavily relies on the approximation that it is only the anisotropy in the total intensity that contributes to the alignment of the molecular or atomic states under investigation. We call this approximation the anisotropic intensity approximation. We were able to directly compare the anisotropic intensity approximation to the LVG problem of \citet{goldreich:81}. \citet{goldreich:81} accounted for the influence of the anisotropy of both the Stokes I and Stokes Q on the quantum state alignment. In the GK approach, the Stokes U component of the radiation field is neglected because the LVG method can only treat a constant magnetic field. The comparison is summarized in Figure \ref{fig:GK_compare}.

We note that below polarization fractions of $2\%$, our method agrees with the GK effect for any optical depth. Furthermore, for low optical depth, $\tau < 0.3$, our method reproduces the GK effect very well regardless of the polarization fraction. It is only for very high degrees of polarization and large optical depths that the polarization fraction obtained through the anisotropic intensity approximation starts to deviate from the GK polarization fraction. For strongly polarized lines ($p_L > 6\%$), the polarization fraction can be underestimated by up to a factor $1.5$ for $\tau > 1$ and this underestimation is sustained with increasing $\tau$. We note that the polarization angle is identical for both methods.

The anisotropic intensity approximation loses its quality through the following: (i) the fact that a significant part of the radiation is polarized, which has an impact on the irreducible tensor representation of the radiation field (see Eq.~\ref{eq:irred_J}), and (ii) that this simplification subsequently impacts the source function, resulting in a magnification of the error. The latter error is particularly manifest at high optical depths, and it is also a consequence of the local approximation of an LVG-like problem. We expect this error to be ameliorated when the local approximation is abandoned as in PORTAL. 

The polarization of (sub)millimeter lines through the GK effect has been observed in a number of sources. For most line observations, the observed polarization fraction is lower than $2\%$ \citep{lai:03}. This can be taken as a direct indicator of the quality of the anisotropic intensity approximation. There is a fraction of emission lines for which high polarization fractions are observed; the most strongly polarized emission lines go up to $13 \%$ \citep{vlemmings:12, cortes:05}. The large polarization fractions are most probably due to large sources of external radiation in the vicinity. 

One avenue to remedy the anisotropic intensity approximation is to iteratively perform the inward ray-tracing steps (see Section \ref{sec:polsee}) for all radiative polarization modes and perform the irreducible tensor integration as Eq.~(\ref{eq:irred_J}). After each iteration, the alignment of the quantum states for each cell is recomputed until convergence is attained. We plan to implement such a scheme in a later version of PORTAL, although this will significantly increase the calculation time.
\subsection{The strong magnetic field approximation}
\label{sec:comp_align}
The symmetry axis of the molecular and atomic states determines the (projected) direction of polarization. In our models, it is assumed that the symmetry axis is along the local magnetic field direction. This requires the magnetic precession rate to be $10$-$100$ times stronger than other directional interaction rates. If an alternative directional interaction is about as strong or stronger than the magnetic precession rate, then the symmetry axis of the quantum states is rotated. 

The magnetic precession rate for a nonparamagnetic molecule is given by 
\begin{align}
g\Omega = 4.8 g_{\mathrm{mol}} \left(\frac{B}{\mathrm{mG}} \right) \ \mathrm{s}^{-1}, 
\end{align} 
where $g_{\mathrm{mol}}$ is the molecular g-factor: A dimensionless factor that determines the coupling of the molecule to the magnetic field. For linear molecules, $g_{\mathrm{mol}}$ is the same for all rotational levels. The molecular g-factors of CO and HCO$^+$ that we consider in this work are $g^{\mathrm{CO}} = -0.269$ \citep{flygare:71} and $g^{\mathrm{HCO}^+}=0.006$.\footnote{
We computed the g-factor of HCO$^+$ using quantum chemical techniques since no experimental data are available. The quantum chemical calculations were performed at the CCSD(T) level of theory, using aug-cc-pVTZ basis sets, with the CFOUR program package \citep{CFOUR}. We used a linear geometry of $r_{\mathrm{CO}}=1.112$ \AA and $r_{\mathrm{CH}}=1.095$ \AA. We note that the molecular g-factor of HCO$^+$ is anomalously low. Indeed, the only polarimetric observation of HCO$^+$ yielded no detection \citep{glenn:97}. This could be an effect of weak Zeeman precession. However, one should not forget that HCO$^+$ has in fact a hyperfine structure, where each hyperfine-transition has its own g-factor \citep[see, for instance,][]{lankhaar:18} that only averages to the rotational g-factor if all hyperfine-transitions have line-strengths proportional to the hyperfine-resolved Einstein A-coefficients. 
}

We compare the magnetic precession rate ($1$ mG and $1$ $\mathrm{\mu G}$) to the cumulative rate of stimulated emission in Figures \ref{fig:sphere_rates} and \ref{fig:pp_rates}. For the problems we considered, the magnetic precession rate is dominant over all other interactions and it is justified to assume that the quantum state symmetry axis is along the magnetic field direction.

Earlier, we saw that HCO$^+$ had an exceptionally low magnetic moment. Conversely, the dipole moment of HCO$^+$ is very large. Thus radiative interactions for such a molecule are very strong, and therefore also a strong magnetic field is required to justify the dominant magnetic field approximation. Indeed, for a large region of the collapsing sphere, a $1 \ \mathrm{\mu G}$ magnetic field would not determine the HCO$^+$ symmetry axis. We stress that for molecules that have strong radiative interactions, one should be extra vigilant and check the relevant interaction rates to verify that the magnetic field truly defines the symmetry axis of the quantum states and thus if the polarization vectors do indeed trace the magnetic field structure. 

It is conceivable that a strong external radiation field that has a large angular size, such as a large stellar object, determines the quantum state symmetry axis. The directional rate of interaction of a general lower quantum state, $1$, by an external black-body radiation source at the solid angle $\Delta \Omega_*$ and with the temperature $T_*$ is \citep{nedoluha:92,morris:85} 
\begin{align}
R_{12} = \frac{g_2}{g_1} A_{21} \left[e^{h\nu_{21} / k_B T_*} - 1 \right]^{-1} \Delta \Omega_*,
\end{align}
where $g_i$ is the degeneracy of level $i$ and $A_{21}$ and $\nu_{21}$ are the Einstein coefficient and frequency of the transition from upper level $2$ to lower level $1$. It is apparent from this expression that (sub)millimeter lines have relatively low interaction rates. Rather, vibrational transitions in the IR region have associated directional interaction rates that are far greater and are more likely to compete with magnetic interactions to determine the symmetry axis of the quantum states. For instance, the interaction rate of the $(v,J)$, $(0,0)\to (1,1)$ transition of CO is $\sim 7.8 \ \mathrm{s}^{-1}$ when it is excited by a $2000$ Kelvin black-body radiation source at $\Delta \Omega_* = 1$ sr. The rate drops quadratically with the distance to the external radiation source and it is not corrected for absorption. We implemented a module in PORTAL that can incorporate the interactions resulting from a bright external source of radiation through vibrational transitions. This is particularly important when investigating the circumstellar envelopes of evolved stars \citep{morris:85, ramos:05}.

The strong magnetic field approximation should be abandoned when multiple directional interactions have similar interaction rates. In that case, one must comprehensively model all anisotropies affecting the quantum state alignment. This can be done at the expense of a computational effort as it increases the dimensionality of the problem greatly. For example, in treating the first $41$ rotational levels of a linear rotor and by setting $k_{\mathrm{max}}=6$, the dimensionality of the SEE would increase from $151$ to $1086$, provided that we neglect orientation elements of uneven $k$. The general theory of setting up the complete SEE can be found in Chapter 7 of \citet{landi:06}. 
\subsection{General remarks}
\label{sec:general_remarks}
\subsubsection{(Sub)millimeter line polarization in astrophysical regions}
It is clear from our calculations that the only requirement for the emergence of polarized emission is a source that has some form of anisotropy. This anisotropy may come from the velocity field, which has already been explored by \citet{goldreich:81}, but it is not necessarily limited to this. To present the capabilities of PORTAL, we computed the emergence of polarized radiation in a protoplanetary disk and a collapsing sphere. In the protoplanetary disk, anisotropy mostly comes from the density structure. For the collapsing sphere, anisotropy comes from both the velocity-field and the density structure.

Furthermore, we confirm the earlier observation of \citet{goldreich:81}, which is that namely around optical depths of unity, the polarization of line emission is the strongest. The physical reason behind this is that for sources with some sort of anisotropy, around $\tau \sim 1,$ this anisotropy is most manifest in the local radiation field. Subsequently leading to the highest polarization degrees. 

\subsubsection{Sampling} 
The sampling of the space that we used is identical to the sampling used by LIME in which a random sampling, weighed by the density-structure, of the space is performed and neighboring cells are found through a Voronoi tessellation. We found that the extensive angular sampling that we performed to compute the local anisotropic radiation field generally requires a higher sampling of the space than would be necessary if one is generating a nonpolarized image. We found that for insufficient sampling of the space, strong local variation in the polarization fractions manifest themselves even though similar variations would not be visible in the total intensity. Also, local $90^o$ flips of the polarization vectors can be a product of sampling of the surrounding space that is too sparse. For a source with symmetry in both the magnetic field and the radiative transfer structure, it can be prudent to use symmetrical averages in the case of a sampling that is too sparse.

\subsubsection{Collisions} 
In order for appreciable polarization in the emission from astrophysical regions to be produced, one requires the rate of (isotropic) collisions to be relatively low. When collisions occur more than $100$ times as frequent as the aligning absorption and stimulated emission events, no observable polarization is produced. Polarization is therefore not produced in regions of high number density and temperature. In general, regions that are in local thermal equilibrium show no appreciable polarization in their emission. 

In the astrophysical problems that we analyzed, we represented collisions only by their rank-0 elements, that is, we assumed all magnetic substates to be equally pumped. At the same time, we assumed no depolarization through elastic collisions. The systematic errors of both assumptions are opposite. Such an approximation for the alignment characteristics of collisional rates is a common assumption in the modeling of alignment of quantum states \citep{landi:06}. Indeed, collisional rates resolved at the level of magnetic substates are not readily available, even though it is possible to compute these using modern quantum-dynamical methods \citep{alexander:79, faure:12, landi:06}. 


\subsubsection{External radiation} 
We found that an external source of directional radiation enhances the polarization appreciably. Similar conclusions have also been drawn in maser polarization theory \citep{lankhaar:19} and also for the GK effect \citep{deguchi:84, cortes:05}. In particular, \citet{cortes:05} found that they could explain the $90^o$-flip in polarization angle between the CO $J=1-0$ and the $J=2-1$ transitions through the anisotropic radiation coming from an external source. We confirm that this is one possible explanation, but we stress that there are other avenues to attain such a polarization effect. According to our theory, this $90^o$-flip is most generally explained by the $\eta_Q^{1-0}$ and the $\eta_Q^{2-1}$ elements being of opposite signs. This does not necessarily require an external radiation source. 

It should be emphasized that polarization enhancement through external radiation is most manifest when hot objects irradiate high-frequency transitions, such as vibrational lines. It is also the case for such transitions that are most likely to compromise the strong magnetic field approximation (see Section \ref{sec:comp_align}). In this work, we have abstained from including higher vibrational states when computing the polarization maps, but we will further explore this when we use PORTAL in conjunction with more detailed models of astrophysical regions and the involved radiative processes. 

\subsubsection{Alternative routes to polarization} 
Dust emission is often observed to be partially polarized. This has been seen in protoplanetary disks \citep{hull:17}, in circumstellar envelopes of evolved stars \citep{vlemmings:17}, and molecular clouds \citep{soler:13}. Polarized emission from dust follows from its alignment. Dust can get aligned to the magnetic field through the process of radiative torque alignment \citep{draine:97}, but alignment to a strong external source of radiation \citep{lazarian:07} or through self-scattering \citep{kataoka:15} is also possible. 

The dust polarization is indicative of the alignment and therefore does not always trace the (projected) magnetic field direction. Polarization fractions are observed to be up to a few percent. In PORTAL, we neglected the contribution of the dust polarization to the molecular state alignment because we used the anisotropic intensity approximation. In the ray-tracing step, we implemented the dust polarization module outlined in \citet{padovani:12} and added it to the regular line polarization ray-tracing. We have found in the simulations we present here that the contribution of the dust polarization around the line-frequency is negligible because the line-opacity is some orders of magnitude greater than the dust opacity. This means that for strong enough magnetic fields (see Section \ref{sec:comp_align}), line polarization faithfully traces the (projected) magnetic field direction with 90$^o$ ambiguity.

Recently, it has been proposed that through forward scattering of radiation by a collective of molecules, a phase difference can be induced to the parallel and perpendicularly polarized components of the radiation field \citep{houde:13}. The phase difference subsequently leads to a conversion of Stokes-U to Stokes-V radiation. This process, called anisotropic resonant scattering, would lead to the production of circular polarization at the cost of linear polarization, and it also changes the polarization angle. Observational evidence for this phenomenon is accruing \citep{hezareh:13, chamma:18}. Anisotropic resonant scattering is typically thought to occur in a foreground cloud, between the observer and the source of polarized line emission \citep{houde:13}, but it could also be a feature of the radiative transfer inside the source. A better estimate of the relative strength of anisotropic resonant scattering has to be developed before we can evaluate the importance of this effect on the emergence of linear polarization in thermal line emission. 

\subsubsection{Ground state alignment.} 
\citet{yan:06} showed that polarization can emerge in atomic (hyper)fine-structure lines through (i) a strong magnetic field that defines the symmetry axis and (ii) an external UV radiation field that induces directional transitions, aligning the quantum states. If the pumping rate is much lower than the spontaneous decay rates of the excited states, only the ground state of the atomic system is aligned. Collisions and stimulated emission events are neglected in the formalism of ground state alignment (GSA). Through neglecting collisions and stimulated emission events and adapting an idealized geometry, \citet{yan:06} are able to formulate semianalytical expressions for the polarization fractions emerging from atomic lines. GSA has been proposed as a polarizing mechanism for atomic lines in the ISM \citet{zhang:18}.

PORTAL builds on the same theory as GSA, but it explicitly incorporates the effect of collisions and stimulated emission events. Furthermore, instead of assuming that a radiation field only comes from an external source, PORTAL maps out the full 3D radiation field structure of the medium in which the investigated species is embedded. In this work, we focus on the polarized radiative transfer of (sub)millimeter molecular and atomic lines because its radiative transfer does not involve any scattering \citep{brinch:10}. We plan to extend our model to also incorporate the emergence of polarization in atomic fine-structure lines, where we will pay special attention to scattering in the radiative transfer of these systems.

\section{Conclusions}
We present PORTAL, a 3D polarized radiative transfer program that is adapted to lines. The program uses the strong magnetic field approximation and the anisotropic intensity approximation, both of which we show to hold for the majority of relevant astrophysical problems. PORTAL can be used in stand-alone mode using an LTE estimate of the molecular or atomic excitation. Alternatively, the output of existing 3D radiative transfer programs can be input in PORTAL. 

To outline PORTAL's capabilities, we computed the polarization maps of a collapsing sphere and a simple protoplanetary disk model. The polarization spectrum of a collapsing sphere shows polarization in its spectral lines up to $2\%$ with the associated polarization vectors aligned with the projected magnetic field direction. The protoplanetary disk when viewed face on shows polarization fractions up to $\sim 0.5 \%$, but the polarization fraction rises to $\sim 9\%$ at significant inclinations. The polarization vectors resulting from a radial and toroidal magnetic field configuration are identical for a face-on view of the protoplanetary disk, and they can only be distinguished when viewed at a significant inclination. In forthcoming papers, we plan to use PORTAL to analyze the emergence of polarization in spectral lines in more detailed models of protoplanetary disks, to a molecular outflow, and to the circumstellar envelopes of AGB stars.

\begin{acknowledgements} Support for this work was provided by the Swedish Research Council (VR). Simulations were performed on resources at the Chalmers Centre for Computational Science and Engineering (C3SE) provided by the Swedish National Infrastructure for Computing (SNIC). Ko-Yun Huang and Athol Kemball are acknowledged for sharing the results of their GK code. The authors thank Luis Velilla Prieto for helpful comments on a first draft of the manuscript. We thank the referee (Martin Houde) for comments that improved the paper.
 \end{acknowledgements}

\bibliography{/Users/boylankhaar/texlibs/lib}

\begin{thebibliography}{47}
\expandafter\ifx\csname natexlab\endcsname\relax\def\natexlab#1{#1}\fi

\bibitem[{Alexander(1979)}]{alexander:79}
Alexander, M.~H. 1979, J. Chem. Phys., 71, 5212

\bibitem[{Anderson {et~al.}(1999)Anderson, Bai, Bischof, Blackford, Demmel,
  Dongarra, Du~Croz, Greenbaum, Hammarling, McKenney, \& Sorensen}]{LAPACK}
Anderson, E., Bai, Z., Bischof, C., {et~al.} 1999, {LAPACK} Users' Guide, 3rd
  edn. (Philadelphia, PA: Society for Industrial and Applied Mathematics)

\bibitem[{Biedenharn {et~al.}(1981)Biedenharn, Louck, \&
  Carruthers}]{biedenharn:81}
Biedenharn, L.~C., Louck, J.~D., \& Carruthers, P.~A. 1981, Angular momentum in
  quantum physics: theory and application (Addison-Wesley Reading, MA)

\bibitem[{Brinch \& Hogerheijde(2010)}]{brinch:10}
Brinch, C. \& Hogerheijde, M. 2010, Astron. Astrophys., 523, A25

\bibitem[{Chamma {et~al.}(2018)Chamma, Houde, Girart, \& Rao}]{chamma:18}
Chamma, M.~A., Houde, M., Girart, J.~M., \& Rao, R. 2018, Mon. Not. R. Astron.
  Soc., 480, 3123

\bibitem[{Cortes {et~al.}(2005)Cortes, Crutcher, \& Watson}]{cortes:05}
Cortes, P.~C., Crutcher, R., \& Watson, W. 2005, Astrophys. J., 628, 780

\bibitem[{Crutcher(2012)}]{crutcher:12}
Crutcher, R.~M. 2012, Annu. Rev. Astron. Astrophys., 50, 29

\bibitem[{Crutcher \& Kemball(2019)}]{crutcher:19}
Crutcher, R.~M. \& Kemball, A.~J. 2019, Front. Astron. Space Sci., 6, 66

\bibitem[{Degl'Innocenti \& Landolfi(2006)}]{landi:06}
Degl'Innocenti, M.~L. \& Landolfi, M. 2006, Polarization in spectral lines,
  Vol. 307 (Springer Science \& Business Media)

\bibitem[{Deguchi \& Watson(1984)}]{deguchi:84}
Deguchi, S. \& Watson, W. 1984, Astrophys. J., 285, 126

\bibitem[{Draine \& Weingartner(1997)}]{draine:97}
Draine, B.~T. \& Weingartner, J.~C. 1997, Astrophys. J., 480, 633

\bibitem[{Faure \& Lique(2012)}]{faure:12}
Faure, A. \& Lique, F. 2012, Mon. Not. R. Astron. Soc., 425, 740

\bibitem[{Flygare \& Benson(1971)}]{flygare:71}
Flygare, W. \& Benson, R. 1971, Mol. Phys., 20, 225

\bibitem[{Glenn {et~al.}(1997)Glenn, Walker, \& Jewell}]{glenn:97}
Glenn, J., Walker, C.~K., \& Jewell, P. 1997, Astrophys. J., 479, 325

\bibitem[{Goldreich \& Kylafis(1981)}]{goldreich:81}
Goldreich, P. \& Kylafis, N.~D. 1981, Astrophys. J., 243, L75

\bibitem[{Goldreich \& Kylafis(1982)}]{goldreich:82}
Goldreich, P. \& Kylafis, N.~D. 1982, Astrophys. J., 253, 606

\bibitem[{Han(2017)}]{han:17}
Han, J. 2017, Annu. Rev. Astron. Astrophys., 55, 111

\bibitem[{Hezareh {et~al.}(2013)Hezareh, Wiesemeyer, Houde, Gusdorf, \&
  Siringo}]{hezareh:13}
Hezareh, T., Wiesemeyer, H., Houde, M., Gusdorf, A., \& Siringo, G. 2013,
  Astron. Astrophys., 558, A45

\bibitem[{Houde {et~al.}(2013)Houde, Hezareh, Jones, \& Rajabi}]{houde:13}
Houde, M., Hezareh, T., Jones, S., \& Rajabi, F. 2013, Astrophys. J., 764, 24

\bibitem[{Hull {et~al.}(2017)Hull, Girart, Tychoniec, Rao, Cort{\'e}s, Pokhrel,
  Zhang, Houde, Dunham, Kristensen, {et~al.}}]{hull:17}
Hull, C.~L., Girart, J.~M., Tychoniec, {\L}., {et~al.} 2017, Astrophys. J.,
  847, 92

\bibitem[{Johansson \& Forssen(2016)}]{johansson:16}
Johansson, H.~T. \& Forssen, C. 2016, SIAM Journal on Scientific Computing, 38,
  A376

\bibitem[{Kataoka {et~al.}(2015)Kataoka, Muto, Momose, Tsukagoshi, Fukagawa,
  Shibai, Hanawa, Murakawa, \& Dullemond}]{kataoka:15}
Kataoka, A., Muto, T., Momose, M., {et~al.} 2015, Astrophys. J., 809, 78

\bibitem[{Kataoka {et~al.}(2017)Kataoka, Tsukagoshi, Pohl, Muto, Nagai,
  Stephens, Tomisaka, \& Momose}]{kataoka:17}
Kataoka, A., Tsukagoshi, T., Pohl, A., {et~al.} 2017, Astrophys. J. Lett., 844,
  L5

\bibitem[{Kuiper {et~al.}(2020)Kuiper, Vlemmings, Girart, Bertoldi,
  J{\o}rgensen, Brinch, Hogerheijde, Juhasz, Padovani, \& Schaaf}]{kuiper:20}
Kuiper, R., Vlemmings, W., Girart, J., {et~al.} 2020, Astron. Astrophys., xxx,
  xxx

\bibitem[{Lai {et~al.}(2003)Lai, Girart, \& Crutcher}]{lai:03}
Lai, S.-P., Girart, J.~M., \& Crutcher, R.~M. 2003, Astrophys. J., 598, 392

\bibitem[{Landi~Degl'Innocenti(1984)}]{landi:84}
Landi~Degl'Innocenti, E. 1984, Sol. Phys., 91, 1

\bibitem[{{Lankhaar} \& {Vlemmings}(2019)}]{lankhaar:19}
{Lankhaar}, B. \& {Vlemmings}, W. 2019, Astron. Astrophys., 628, A14

\bibitem[{Lankhaar {et~al.}(2018)Lankhaar, Vlemmings, Surcis, van Langevelde,
  Groenenboom, \& van~der Avoird}]{lankhaar:18}
Lankhaar, B., Vlemmings, W., Surcis, G., {et~al.} 2018, Nat. Astron., 2, 145

\bibitem[{Lazarian \& Hoang(2007)}]{lazarian:07}
Lazarian, A. \& Hoang, T. 2007, Mon. Not. R. Astron. Soc., 378, 910

\bibitem[{Morris {et~al.}(1985)Morris, Lucas, \& Omont}]{morris:85}
Morris, M., Lucas, R., \& Omont, A. 1985, Astron. Astrophys., 142, 107

\bibitem[{Nedoluha \& Watson(1992)}]{nedoluha:92}
Nedoluha, G.~E. \& Watson, W.~D. 1992, Astrophys. J., 384, 185

\bibitem[{Padovani {et~al.}(2012)Padovani, Brinch, Girart, J{\o}rgensen, Frau,
  Hennebelle, Kuiper, Vlemmings, Bertoldi, Hogerheijde, {et~al.}}]{padovani:12}
Padovani, M., Brinch, C., Girart, J., {et~al.} 2012, Astron. Astrophys., 543,
  A16

\bibitem[{Ramos {et~al.}(2005)Ramos, Degl’Innocenti, \& Bueno}]{ramos:05}
Ramos, A.~A., Degl’Innocenti, E.~L., \& Bueno, J.~T. 2005, Astrophys. J.,
  625, 985

\bibitem[{Ritzerveld \& Icke(2006)}]{ritzerveld:06}
Ritzerveld, J. \& Icke, V. 2006, Phys. Rev. E, 74, 026704

\bibitem[{Rybicki \& Hummer(1991)}]{rybicki:91}
Rybicki, G.~B. \& Hummer, D.~G. 1991, Astron. Astrophys., 245, 171

\bibitem[{Shu(1977)}]{shu:77}
Shu, F.~H. 1977, Astrophys. J., 214, 488

\bibitem[{Soler {et~al.}(2013)Soler, Hennebelle, Martin, Miville-Desch{\^e}nes,
  Netterfield, \& Fissel}]{soler:13}
Soler, J.~D., Hennebelle, P., Martin, P., {et~al.} 2013, Astrophys. J., 774,
  128

\bibitem[{Stanton {et~al.}(2009)Stanton, Gauss, M.E., \& Szalay}]{CFOUR}
Stanton, J., Gauss, J., M.E., H., \& Szalay, P. 2009, {CFOUR},
  {C}oupled-{C}luster techniques for {C}omputational {C}hemistry,
  http://www.cfour.de

\bibitem[{Stephens {et~al.}(2017)Stephens, Yang, Li, Looney, Kataoka, Kwon,
  Fern{\'a}ndez-L{\'o}pez, Hull, Hughes, Segura-Cox, {et~al.}}]{stephens:17}
Stephens, I.~W., Yang, H., Li, Z.-Y., {et~al.} 2017, Astrophys. J., 851, 55

\bibitem[{van~der Tak {et~al.}(2007)van~der Tak, Black, Sch{\"o}ier, Jansen, \&
  van Dishoeck}]{vandertak:07}
van~der Tak, F., Black, J.~H., Sch{\"o}ier, F., Jansen, D., \& van Dishoeck,
  E.~F. 2007, Astron. Astrophys., 468, 627

\bibitem[{{van Zadelhoff} {et~al.}(2002){van Zadelhoff}, {Dullemond}, {van der
  Tak}, {Yates}, {Doty}, {Ossenkopf}, {Hogerheijde}, {Juvela}, {Wiesemeyer}, \&
  {Sch{\"o}ier}}]{zadelhoff:02}
{van Zadelhoff}, G.-J., {Dullemond}, C.~P., {van der Tak}, F.~F.~S., {et~al.}
  2002, Astron. Astrophys., 395, 373

\bibitem[{Vlemmings(2013)}]{vlemmings:13}
Vlemmings, W. 2013, Proceedings of the International Astronomical Union, 9, 389

\bibitem[{Vlemmings {et~al.}(2017)Vlemmings, Khouri, Mart{\'\i}-Vidal, Tafoya,
  Baudry, Etoka, Humphreys, Jones, Kemball, O’Gorman,
  {et~al.}}]{vlemmings:17}
Vlemmings, W., Khouri, T., Mart{\'\i}-Vidal, I., {et~al.} 2017, Astron.
  Astrophys., 603, A92

\bibitem[{Vlemmings {et~al.}(2019)Vlemmings, Lankhaar, Cazzoletti, Ceccobello,
  Dall’Olio, van Dishoeck, Facchini, Humphreys, Persson, Testi, \&
  Williams}]{vlemmings:19}
Vlemmings, W., Lankhaar, B., Cazzoletti, P., {et~al.} 2019, Astron. Astrophys.,
  624, L7

\bibitem[{Vlemmings {et~al.}(2012)Vlemmings, Ramstedt, Rao, \&
  Maercker}]{vlemmings:12}
Vlemmings, W., Ramstedt, S., Rao, R., \& Maercker, M. 2012, Astron. Astrophys.,
  540, L3

\bibitem[{Yan \& Lazarian(2006)}]{yan:06}
Yan, H. \& Lazarian, A. 2006, Astrophys. J., 653, 1292

\bibitem[{Zhang \& Yan(2018)}]{zhang:18}
Zhang, H. \& Yan, H. 2018, Mon. Not. R. Astron. Soc., 475, 2415

\end{thebibliography}

\end{document}